\documentclass[twocolumn]{aa}  

\newcommand{\lya}{Ly$\alpha$}

\newcommand{\bet}{$\beta$}

\newcommand{\ebv}{$E_{B-V}$}

\newcommand{\flya}{$F_{{\rm Ly}\alpha}$}
\newcommand{\wlya}{$W_{{\rm Ly}\alpha}$}

\newcommand{\kms}{\mbox{km~s$^{-1}$}}

\usepackage{graphicx}
\usepackage{txfonts}
\begin{document}
\title{On the narrowband detection properties of high-redshift Lyman-alpha emitters}

\author{Matthew Hayes \and G\"oran \"Ostlin}

\offprints{M. Hayes}

\institute{Stockholm Observatory, AlbaNova University Centre,
SE-106 91 Stockholm, Sweden\\
\email{matthew@astro.su.se}
}
\date{Received April 25 2006; accepted September 05, 2006}


  \abstract
{Numerous surveys are currently underway or planned that aim to exploit
the strengths of the Lyman-alpha emission line for cosmological purposes.  
Today, narrowband imaging surveys are frequently used as a probe of the 
distant universe.}
{To investigate the reliability of the results of such high-$z$ \lya\ 
studies, and the validity of the conclusions that are based upon
them. 
To determine whether reliable \lya\ fluxes (\flya) and equivalent widths
(\wlya) can be estimated from narrowband imaging surveys and whether any
observational biases may be present.}
{We have developed software to simulate the observed line and
continuum properties of synthetic \lya\ galaxies in the distant universe by
adopting various typical observational survey techniques. 
This was used to investigate how detected \flya\ and \wlya\  
vary with properties of the host galaxy or intergalactic medium: internal dust
reddening; intervening \lya\ absorption systems; the presence of
underlying stellar populations.}
{None of the techniques studied are greatly susceptible
to underlying stellar populations or the relative contribution of nebular 
gas.
We find that techniques that use one off-line filter on the red
side of \lya\ result in highly inaccurate measurements of \wlya\ under
all tests.
Adopting two off-line filters to estimate continuum at \lya\ is an
improvement but is still unreliable when dust extinction is considered.
Techniques that employ single narrow- and broad-band filters with the same
central wavelength are not susceptible to internal dust, but \lya\ 
absorption in the IGM can cause \wlya\ to be overestimated by factors of up 
to 2: at $z=6$, the median \wlya\ is overestimated by $\sim25$\%. 
The most robust approach is a SED fitting technique that 
fits \ebv\ and burst-age from synthetic models --
broadband observations are needed that sample the UV continuum slope, 2175\AA\
dust feature, and the 4000\AA\ discontinuity. 
We also notice a redshift-dependent incompleteness that results from DLA
systems close to the target LAEs, amounting to $\sim 10$\% at $z=6$.
}
{}


\titlerunning{The detection properties of high-$z$ \lya\ galaxies}
\authorrunning{Hayes \& \"Ostlin}
\maketitle

%
\section{Introduction\label{sect:intro}}

Galaxies hosting young, violent starbursts are expected to accommodate
numerous massive, hot stars that are bright in the far ultraviolet (UV)
regime.
UV photons with wavelengths shortwards of the Lyman break ionise their
local interstellar media (ISM), resulting in recombination
nebulae bright in hydrogen lines. 
Two-thirds of the Lyman-continuum photons are reprocessed as Lyman-alpha
(\lya) under the assumption of case B recombination, and it could be
expected that young starburst regions should be consistently
\lya-bright.

Early observations of local star-forming galaxies with the {\em 
International Ultraviolet Explorer} (IUE) demonstrated this not to be 
the case. 
In small samples of relatively un-evolved galaxies (low dust and metal
content) \lya\ was frequently shown to be absent or unexpectedly weak
(eg. 
 Meier \& Terlevich \cite{ref:meier82};
 Hartmann et al. \cite{ref:hartmann88};
 Calzetti \& Kinney \cite{ref:calkin92}).
Although dust could be invoked to explain the absence of
this FUV emission line, 
Giavalisco et al. (\cite{ref:giavalisco96}) 
demonstrated that pure dust
extinction could not explain the weak \lya\ fluxes from any of the
galaxies in the {\em IUE} samples. 
Instead results suggest that \lya\ photons have 
decoupled from the non-resonant continuum radiation, resulting in
line-photons escaping the host over significantly extended 
path-lengths; thereby increasing their sensitivity to dust 
(Neufeld \cite{ref:neufeld90}).
On the other hand, Neufeld (\cite{ref:neufeld91}) 
also found that resonance scattering can cause {\em less} attenuation of
\lya\ photons compared to continuum if the ISM is multiphase: 
if dusty, neutral clouds are embedded in a dust-free, ionised ISM,
resonance scattering can reflect \lya\ photons from the surface of the
clouds, allowing them to diffuse out of the ISM relatively unattenuated.
Radiative transport in such multiphase, clumpy media has more recently 
been studied by 
Hansen \& Oh (\cite{ref:hansen05}) 
who found \lya\ equivalent widths (\wlya) could be `boosted' by factors 
of $2-3$.

In the local universe, the observational situation was 
further-complicated by \lya\ spectroscopic observations using the
{\em Goddard High-Resolution Spectrograph (GHRS)} and {\em Space 
Telescope Imaging Spectrograph (STIS)} both onboard {\em Hubble Space 
Telescope (HST)}. 
In a sample of 8 local starbursts, 
Kunth et al. 
(\cite{ref:kunth98}) 
found that when \lya\ is seen in emission, blueshifted \lya\ absorption
features are also common, leading to P-Cygni-like profiles. 
In the same study, \ion{O}{i} and \ion{Si}{ii} UV absorption lines were 
also blueshifted with respect to the ionised gas, indicating large-scale
outflows of the ISM with velocities up to 200~\kms.
Mas-Hesse et al. 
(\cite{ref:mas-hesse03})
demonstrated that varying \lya\ profiles from starbursts are well
explained by an evolutionary super-shell model; over the lifetime of a
starburst, absorption, pure emission, and P-Cygni emission phases will
all be observed, depending on the properties of the outflow and viewing
angle.
If \lya\ escape is associated with winds and photons can diffuse
through \ion{H}{i}, low surface brightness \lya\ emission may be expected 
to span large areas, significantly more extended than the starburst 
itself as probed by FUV emission. 
Hence imaging becomes an efficient method by which the line can be
isolated.
This was the motivation for our HST survey using the {\em Advanced
Camera for Surveys} (ACS;
Kunth et al. \cite{ref:kunth03}).
Such \lya\ imaging studies are extremely sensitive to how the continuum 
is subtracted and much supporting data is required in order to estimate
the continuum flux at \lya.
We found in 
Hayes et al. (\cite{ref:hayes05}) 
that at least three off-line imaging observations are necessary in order
to estimate the continuum flux at \lya.
We determined it necessary to sample the UV continuum slope and 4000\AA\ break
in order to disentangle the effects of age and reddening, allowing us to  
estimate the flux due to continuum processes in the on-line filter from 
synthetic models. 
In our sample, luminous blue compact galaxy ESO\,338-IG04 shows regions 
of diffuse \lya\ emission spanning several kpc, and patchy structure 
consistent with an inhomogeneous and outflowing ISM. 
In the diffuse regions \wlya\ is shown to be very large $(>200\AA)$
where continuum flux is weak, but integrated over the whole galaxy
\wlya\ is $\sim20$\AA, an order of magnitude lower.
Damped absorption with \wlya$\sim -40$\AA\ is also observed in some
central regions, demonstrating the sensitivity of \lya\ to 
small-scale variations in the ISM.

Due to its sensitivity to star-forming activity, high natural
equivalent widths (up to 240\AA; 
Charlot \& Fall \cite{ref:cf93})
sensitivity to dust, and convenient spectral positioning in the FUV, 
\lya\ has long been considered a competitive tracer of primeval high-$z$
galaxies as they form their first generation of stars.
Partridge \& Peebles 
(\cite{ref:pp67})
suggested that in a collapsing \ion{H}{i} halo, as much as 7\% of the
gravitational energy could be radiated in the \lya\ line.  
While surveys that target continuum emission or utilise dropout techniques may
have been highly successful in recent years, they all share one common
weakness: a luminosity bias in favour of galaxies with strong continuum. 
Such a bias leads to incompleteness at low luminosity 
and these surveys would miss the numerous low-mass, star-forming
objects; the dominant species in the paradigm of hierarchal galaxy 
formation.

The early days of high-$z$ \lya\ studies were largely unsuccessful (see 
Pritchet 
\cite{ref:prit94} 
for a review) and it was not until the late 1990's that searches for 
high-$z$ \lya-emitters (LAE) became fruitful.
Despite the complications, high-$z$ \lya-emitting objects are now routinely 
being discovered by imaging techniques 
(Cowie \& Hu \cite{ref:cowiehu98}; 
Rhoads et al. \cite{ref:rhoads00}; 
Fynbo et al. \cite{ref:fynbo03}; 
Kodaira et al. \cite{ref:kodaira03}; 
Yamada et al. \cite{ref:yamada05}),  
fields surrounding massive objects 
(eg. Monaco et al. \cite{ref:monaco05}), and 
spectroscopic observations
(eg. Kurk et al. \cite{ref:kurk04}). 
After much initial disagreement, the predicted and observed luminosity
functions (LF) of LAEs at high-$z$ may be converging
(Le Delliou et al. \cite{ref:ledelliou05}). 
Interestingly however, a fraction of the LAEs uncovered in the high-$z$
universe have very large measured equivalent widths 
(Cowie \& Hu \cite{ref:cowiehu98}; 
Malhotra \& Rhoads \cite{ref:malrhoads02}). 
In the case of Malhotra \& Rhoads (\cite{ref:malrhoads02}), 
their LAE population at $z=4.5$ was shown to have a median \wlya\ of
400\AA, 65\% greater than the maximum value of 240\AA\ 
(Charlot \& Fall, \cite{ref:cf93})
that can 
be generated by pure star-formation models with normal metallicities 
and IMFs. 
These authors interpreted this to imply that either these objects were
type II quasars or star-forming galaxies with very top-heavy IMFs or
zero-metallicity stars. 
Subsequent X-ray observations with the Chandra satellite ruled out the
possibility that these objects could be AGN 
(Wang et al. \cite{ref:wang04}).
Jimenez \& Haiman (\cite{ref:jimenez06}) 
demonstrate how this (and other effects) can be explained if 10-30\% of
the stars in these galaxies are primordial. 
`Normal' starburst conditions could still explain such high equivalent
widths but the system would have to be perfectly configured so 
as to allow \lya\ photons to leak out while dust blocks much of the
stellar UV radiation. 
Another possibility is that this is the result of some other
astrophysical effect, external to the host galaxy that may cause certain
observational techniques to overestimate \wlya\ when such narrowband
imaging techniques are used. 

While \lya\ may provide a `clean' probe of the high-$z$
universe, the common observational trade-off is evident: spectroscopic
studies may be rich in information regarding the line, while narrowband
imaging studies may be efficient for detections. 
Future generations of high-$z$-optimised integral field units (eg. the  
{\em Multi Unit Spectroscopic Explorer (MUSE)} bound for ESO's VLT) have
the potential to provide significant advances in this area. 
However, in the cases where narrowband imaging alone is used to derive
information about the line, it is essential to examine exactly how
efficient the technique is in doing so. 
To our knowledge, no such comprehensive study has previously been 
performed and this article represents the first step in doing so.
We here simulate how various survey techniques may estimate \lya\
detection properties -- line flux (\flya) and equivalent width (\wlya) -- 
of galaxies at high-$z$.

The paper is organised as follows: 
in Sect. \ref{sect:sw} we describe the software; 
in Sect. \ref{sect:sim} we present the simulations we have performed; 
in Sect. \ref{sect:resdis} we present and discuss some of the 
more important results; and 
in Sect. \ref{sect:conc} we present our concluding remarks.
We assume a flat cosmology with 
$\Omega_\mathrm{M}=0.3$, 
$\Omega_\Lambda=0.7$, 
$H_0=70$~km~s$^{-1}$~Mpc$^{-1}$ throughout.

%
\section{The software\label{sect:sw}}

The \lya\ galaxy simulation software follows a simple three-step 
prescription: 
(i) the restframe spectral energy distribution (SED) of a the 
test galaxy is generated and the \lya\ line is added; 
(ii) the SED is redshifted and the effects of intervening material (IGM
absorption) are applied to the spectrum; and 
(iii) the spectrum is convolved with various filter response profiles
and fluxes are computed. 

\subsection{SED  generation\label{sect:sedgen}}

Restframe SEDs of starburst galaxies are generated from the 
{\em Starburst99} (hereafter SB99; 
Leitherer et al.  \cite{ref:leitherer99}; 
V\'azquez \& Leitherer \cite{ref:vazlei04}) 
synthetic spectral models.
No nebular emission lines are thought to be strong enough to significantly 
contribute to fluxes in the UV filters considered here (see Sect. 
\ref{sect:fluxcomp} for a description of the filters) --
filters are broad and continuum dominated, and in Hayes
et al. (\cite{ref:hayes05}) we concluded that optical emission lines had
negligible impact upon our study. 
Hence the SB99 spectra are used unmodified, free of nebular emission 
lines but including nebular continuum emission.
The software was designed for flexibility and to cover as wide 
parameter space as possible, enabling the user to perform a wide array
of parameter dependency tests. 
Spectra can be selected from the 1999 SB99 data-release, choosing
with burst modes (instantaneous or continuous), 
composition (stellar-only or stellar+nebular), metallicity, initial mass 
function, and age. 
In addition to the standard set of spectra, SED modification code 
was written to allow the selection of:

\begin{enumerate}

\item Departures from the single stellar population (SSP).
It is well known from studies of the low- and high-$z$ universe 
that star-forming galaxies often exhibit composite spectra, with
contributions from the current starburst and an underlying, older 
stellar population. 
Our method for testing the effects of such population mixing is to
allow any single SB99 spectrum to act as a template for a secondary
population. 
This secondary spectrum is renormalised so it contributes a specified 
fraction of the spectrum under consideration at a user-specified 
wavelength, and is added to the spectrum or set of spectra. 

\item Variations in the contribution of nebular gas.
By subtracting out the nebular component from corresponding 
pairs of spectra, the nebular gas spectrum as a function of age
is recovered.
The gas spectrum can then be re-scaled and added back to the 
stellar-only component.
Thus a normalisation factor of 0 corresponds to a stellar-only spectrum,
1 corresponds to the standard stellar+nebular spectrum, and factors $>1$
imply the gas spectrum has been boosted. 

\end{enumerate}

Restframe spectra are rebinned to 1\AA\ steps in wavelength. 
After generation of the template spectrum, the whole SED is renormalised 
to $L_{\lambda}$ at 1500\AA\ based upon input star-formation rate 
($SFR$) using the relationship 
\begin{equation}
L_{\lambda}(1500\mbox{\AA}) = 
 9.5238 \times 10^{39} SFR(\mbox{M}_{\sun} \mbox{yr}^{-1})
\end{equation}
adapted from Kennicutt (\cite{ref:kenn98}).
The all-important \lya\ line of specified equivalent width is 
added to the spectrum at wavelength of 1216\AA.  
At the rest wavelength of \lya, 1\AA\ corresponds to the velocity
of 250~km~s$^{-1}$. 
In addition, pure imaging studies aren't sensitive to line profiles and our
aim is simply to investigate how well intrinsic line-strengths are
recovered (whatever their origin or profile).
Hence all the energy is deposited in one 1\AA\ bin.
See Sect. \ref{sect:fluxcomp} for a more detailed discussion of this.  

\lya\ is the only Lyman series feature added to the SED -- the
\ion{O}{vi}--Ly$\beta$--\ion{C}{ii} complex between 1032 and 1038\AA\
does not fall within any of the filter bandpasses considered here (Sect.
\ref{sect:fluxcomp}).
The \ion{C}{iv}$\lambda$1549\AA\ absorption feature (although
frequently observed with P-Cygni profiles), common in star-forming
galaxies, does fall centrally in one of the bandpasses we use. 
In a sample of 45 local galaxies observed with the {\em IUE}, 
Heckman et al. (\cite{ref:heckman98}) 
find a (\ion{C}{iv}+\ion{Si}{iv})/2 median equivalent width of
-4.6\AA, with the
absorption features never stronger than -10.1\AA. 
Crowther et al. (\cite{ref:crowther06}) 
re-measured these equivalent widths, finding an offset of
1-2\AA\ relative to the Heckman et al. result. 
The filter bandpass that the \ion{C}{iv} feature falls in is broad 
($(1+z)\cdot 300$\AA, centred at 1500\AA; 
see Sect. \ref{sect:fluxcomp}) so the presence of the strongest locally
observed \ion{C}{iv} feature would affect the integrated flux by only 
$\sim3$\% with no redshift dependence. 
Moreover, these authors found that this equivalent width is positively 
correlated with metallicity. 
Since strong \lya\ emission is often interpreted as a sign of {\em
low-metallicity} stars, we decided not to apply the \ion{C}{iv} feature
to our SEDs.

In order to complete the restframe SED, it can be reddened using the
Galactic laws of Seaton (\cite{ref:seaton79}), or Cardelli et al.
(\cite{ref:ccm89}), the SMC law fit of Pr\'evot et al. (\cite{ref:prevot84}), 
the LMC law of Fitzpatrick et al. (\cite{ref:fitz85}), and the 
Starburst law of Calzetti et al. (\cite{ref:calzetti00}).

\subsection{Effects of cosmic distance\label{sect:sedshift}}

The restframe SED is first shifted to the desired redshift and the 
luminosity distance $d_\mathrm{L}$ computed from the formula
\begin{equation}
d_\mathrm{L} = 
 (1+z) \frac{c}{H_0} \int_0^{z} 
 \frac{dz^\prime}{\sqrt{\Omega_\mathrm{M}(1+z^\prime)^3 + \Omega_\Lambda}}
\end{equation}
where $H_0$ is the Hubble constant and $\Omega_\mathrm{M}$ and 
$\Omega_\Lambda$ are
the energy densities of total matter and cosmological constant, 
respectively.
The luminosity at each wavelength is then converted to flux using the
inverse square law. 

Given that we want to study all the possible astrophysical effects,
including possible extreme cases, we chose
to treat the effects of intervening \ion{H}{i} clouds by generating
random distributions of individual clouds, not by applying some 
general average prescription (eg. Madau \cite{ref:madau95}). 
The effect of intervening \ion{H}{i} clouds (\lya\ forest (LAF) to 
damped \lya\ (DLA) systems) is implemented by first assuming the number 
density of absorbing clouds
as a function of redshift takes the form of the power-law 
$N(z) = N_0 (1+z)^{\gamma}$ in the redshift range 0 to $z$ 
where $N_0 = 0.07^{+0.13}_{-0.04}$ and 
$\gamma = 2.45^{+0.75}_{-0.65}$ (Peroux et al. \cite{ref:peroux03}).
Firstly the total number of clouds in the given redshift range is 
generated by biasing a uniform variate pseudo-random number (PRN) 
by the distribution function, within the observational constraints. 
For each cloud, a position in redshift space is generated in the 
same manner: by weighting a PRN by the distribution function. 
The column density $(n_{\mathrm{HI}})$ of each cloud is generated by 
assuming
the distribution obeys the power-law $N(n_{\mathrm{HI}}) \propto
	n_{\mathrm{HI}} ^ {-\beta}$ 
(Rao et al. \cite{ref:rao05})
for $\beta=1.4\pm0.2$ and column densities in the range 
$13 \le \log (n_{\mathrm{HI}} [\mathrm{cm^{-2}]}) \le 22$. 

Once the redshift--column-density distribution is generated, the effect
on the SED of each cloud is determined from its equivalent width
in absorption ($W_\mathrm{abs}$). 
We compute restframe $W_\mathrm{abs}$ from the effective optical depth
at line-centre  ($\tau_0$) following the curve of growth method
(Spitzer \cite{ref:spitzer78}). 
After cosmological scaling $(1+z_{\mathrm{cloud}})$, 
$W_\mathrm{abs}$ and the continuum flux-density at
$(1+z_{\mathrm{cloud}})\cdot \lambda_{\mathrm{Ly\alpha,0}}$  
are then used to
remove the requisite amount of flux from the appropriate bins in the
redshifted spectrum.
In cases where the absorption is damped ($W_\mathrm{abs}$ is greater 
than the bin size), the flux in that bin is set to zero and flux is
symmetrically removed from neighbouring bins until the required
$W_\mathrm{abs}$ has been met.  

Significant attention has been paid to the possibility of
reddening of high-$z$ quasar spectra by DLA systems (eg. 
Pei, Fall \& Bechtold \cite{ref:pei91}).
However, recent studies have demonstrated the typical mean cumulative 
reddening from DLA systems to be small.
For example 
Ellison, Hall \& Lira (\cite{ref:ellison:05}) 
found mean $E_{B-V} < 0.04$ at the $3\sigma$ level for SMC-type dust in a 
sample of 14 high-$z$ quasars with DLAs, while from a sample of 72 DLA 
quasars from the {\em Sloan Digital Sky Survey},
Murphy \& Liske (\cite{ref:murphy04}) 
found mean $E_{B-V} < 0.02$ at $3\sigma$ for SMC dust.
On the other hand, 
Wild et al. (\cite{ref:wild06}) 
find evidence for significant reddening in \ion{Ca}{ii} and
\ion{Mg}{ii} absorbing systems: $E_{B-V} = 0.105$ for SMC type dust in 
the selected strongest absorbers ($E_{B-V} = 0.066$ for their complete
sample). 
Such objects are shown to be similar to DLAs with a number density of
$\sim 20$--30\% that of DLAs at the same redshift. 
While we do not claim that some high-$z$ LAEs would not be heavily dust
reddened,  we chose for this part of the study not to implement
dust-reddening by high-\ion{H}{i} column-density systems.

At $z=0$, the redshifted  SED can be reddened once more using the laws of 
Cardelli et al. (\cite{ref:ccm89}) 
or 
Seaton (\cite{ref:seaton79}) 
to simulate Milky-Way reddening. 


\subsection{Filters and computation of line fluxes and equivalent widths 
	\label{sect:fluxcomp}}

For the sake of simplicity (our aim was to study astrophysical
processes, not observational complexities) the filters are perfect. 
Defined only by their central wavelength and effective width, they 
provide 100\% transmission inside the passband and zero otherwise.
That is, the filters are as close to perfect top-hats as possible,
without introducing numerical, resolution-dependent inconsistencies 
at the edges. 
{\em Narrowband} filters targeting the \lya\ line are defined to have 
a full width of
2\% their central wavelength $(\lambda_\mathrm{c})$ while 
{\em broadband} filters have full widths of 20\%
$\lambda_\mathrm{c}$.
That is, the width of the broadband filters scales with
$\lambda_\mathrm{c}$ in approximately the same way as it does in the
Johnson-Cousins/Bessell and {\em JHKLM} systems. The narrowband filters
scale similarly so as not to introduce effects that may result from
sampling different spectral regions at different redshifts. 

One issue that arises when using narrow filters to isolate an emission
line is the positioning of the line within the filter profile.
That is, is the line centrally positioned and narrow enough to allow
maximum transmission of the line flux?
Chemistry and manufacturing processes dictate that filters (particularly 
narrowband)
cannot be perfectly rectangular and most frequently take the form of
bell-curves. 
Whilst photometric redshifts or drop-out techniques may be used to
remove interlopers (most notably [\ion{O}{ii}]$\lambda3726$\AA),
photo-$z$'s of individual objects do not reach the required levels of 
accuracy to determine the position of \lya\ within a narrow bandpass
(typical photo-$z$ accuracies of $\Delta z=0.05(1+z)$ cannot determine 
whether the line falls within a 2\%$\lambda_\mathrm{c}$ narrowband filter 
or not). 
In addition, photo-$z$ methods tend to be reliant upon sampling the
Balmer/4000\AA\ break which is lost from optical multiband datasets at
$z\sim 1$, something we discuss in another context in Sect.
\ref{sect:resdis}. 
At high-$z$ and without spectroscopic data, the selection of LAE
candidates by detection in the narrowband filter becomes the most 
accurate estimate of redshift. 
Hence effects that result from the line falling in the wing of a
filter will always be something that has to be considered.
Starburst galaxies do not exhibit significantly broadened spectral
lines: concerning \lya, 
Matsuda et al.
(\cite{ref:matsuda06})
spectroscopically measured the widths of 37 LABs at $z=3.1$, finding 
\lya\ full widths (FWHM) in the range 150-1700~km~s$^{-1}$ with a median
value of 550~km~s$^{-1}$. 
This corresponds to a maximum Gaussian $\sigma$ of 3\AA\ at \lya\
(median value of $\sim 1$\AA). 
Thus given the 2\%$\lambda_\mathrm{c}$ narrowband filters used here,
even the broadest \lya\ lines will be completely transmitted (2\%
rectangular filters drop to 99\% of complete line transmission at
$\sigma=4.70\AA\ \sim FWHM = 2700$~km~s$^{-1}$).
The bin size of our SEDs is 1\AA\ and, since the median $\sigma$ of the
Matsuda et al. study is 1\AA, we feel safe in depositing all of the \lya\
flux in one bin (see Sect. \ref{sect:sedgen}). 

`Observed' fluxes in all filters are computed by convolution of the
profile and the spectrum, and integration.
In order to estimate true line-fluxes (\flya), it is essential to have 
a robust estimate of the flux due to continuum processes that is present 
in the on-line filter. 
This estimate of the continuum flux at 1216\AA\ is central to
\lya\ studies and here we use various methods to scale the nearby
continuum flux to \lya\ (described below).
With estimates of the line and continuum fluxes, equivalent widths,
defined as $W_\mathrm{Ly\alpha} = F_\mathrm{Ly\alpha} / f_{\mathrm{cont}}$,
are computed, where $f_{\mathrm{cont}}$ is the continuum flux density at
line-centre.
To this end, a number of different commonly implemented and feasible 
observational techniques are employed from which \flya\ and \wlya\ are 
estimated.

\begin{figure}[htbp]
\centering
\includegraphics[width=7cm]{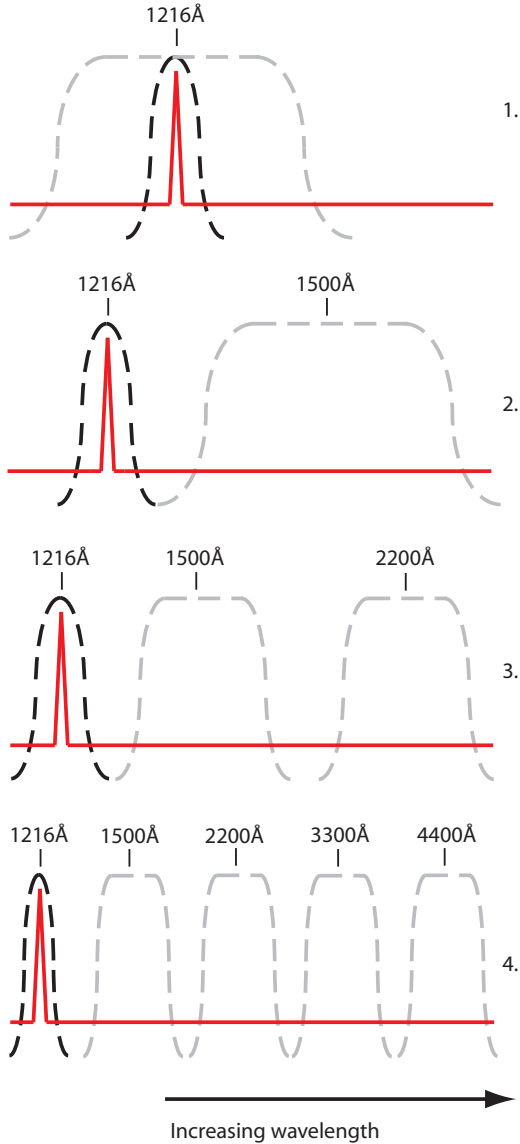}
\caption{Schematic diagram of filter positioning for the four 
observational approaches described in Sect. \ref{sect:fluxcomp} the text. 
Narrowband filters (on-line, black) have widths of 2\% their central
wavelengths and 
broadband filters (continuum, grey) have widths of 20\%
$\lambda_\mathrm{c}$. 
Wavelength indications represent central wavelength of the filters and
are marked in the restframe of the SED. I.e. observation-frame filters 
are centred at $(1+z) \cdot \lambda_\mathrm{c}$.
[{\em See the electronic journal article for the colour version of this figure.}]
}
\label{fig:filters}
\end{figure}

For on-line observations, a narrowband (2\%$\lambda_\mathrm{c}$) filter is 
used, centred at the observed wavelength of redshifted \lya. 
Broadband (20\%$\lambda_\mathrm{c}$) filters are positioned at various 
wavelengths in the
redshifted SED in the UV and optical, sampling the restframe SED at 
1216, 1500, 2200, 3300, and 4400\AA.
That is, filter wavelengths and full widths are defined in the
restframe -- the same region of the SED is sampled at all redshifts.
Using these broadband filters, continuum flux at \lya\ is estimated
using four techniques numbered (\#1-\#4) in order of the central
wavelength of the reddest filter used. A schematic 
representation can be seen in Fig. \ref{fig:filters}.

The techniques are: 
\begin{enumerate}
\item Narrowband on-line filter at redshifted \lya\ and 
{\em one} broadband filter with the same central wavelength. 
While neither of these observations samples the line or continuum 
individually, they both sample the line-flux plus continuum; the flux in
each filter is the sum of the line-flux (\flya) and the product of the
continuum flux-density ($f_{cont}$) and filter-width ($W$). 
If a flat continuum $(\beta=0)$ is assumed across the broadband filter, 
\flya\ and  $f_{cont}$ can be obtained from the solution of
two simultaneous equations as 

\begin{equation}
F_{line} = \frac{W_\mathrm{b}F_\mathrm{n} -
	W_\mathrm{n}F_\mathrm{b}}{W_\mathrm{b} - W_\mathrm{n}} \\ 
	f_{cont} = \frac{F_\mathrm{b} - F_\mathrm{n}}{W_\mathrm{b} - W_\mathrm{n}}
\label{eq:fluxew}
\end{equation}

where $F$ and $W$ represent the flux and filter width, and subscripts 
$n$ and $b$ refer to the broad and narrowband filters, respectively. 
This is the same observational setup as used by Cowie \& Hu
(\cite{ref:cowiehu98}) 
and
Fynbo et al. (\cite{ref:fynbo02}), 
and dividing the first part of Equ. \ref{eq:fluxew} by the second, 
one obtains the equivalent width expression used by Malhotra \& Rhoads 
(\cite{ref:malrhoads02}).

\item Narrowband on-line filter centred on redshifted \lya\ and 
{\em one} completely off-line broadband filter centred at 
$(1+z) \cdot 1500$\AA\ to sample the continuum. 
At 1500\AA, a 20\% (300\AA\ full width) filter will not transmit the
\lya\ line itself.
The continuum flux at \lya\ can then be estimated by  assuming 
the continuum can be approximated by a power-law of the form 
$f_{\mathrm{\lambda}} \propto \mathrm{\lambda}^{\beta}$, with arbitrary
index \bet. 
This technique allows the investigation of the reliability of assuming
the behaviour of the UV continuum slope in the situation where only one 
continuum observation is available.

\item Narrowband on-line filter centred at redshifted \lya\ and {\em two}
broadband continuum filters centred at $(1+z) \cdot 1500$ and 
$(1+z) \cdot 2200$\AA.
This method assumes the UV continuum can be approximated by a 
power-law of the same form as in technique \#2. 
From the two continuum fluxes, the index \bet\ is measured between
restframe 1500\AA\ and 2200\AA. 
This \bet\ is then used to scale the 1500\AA\ flux to the domain sampled
by the narrowband filter and estimate $f_\mathrm{cont}$ at 1216\AA.

\item Narrowband on-line filter centred at redshifted \lya\ 
and {\em four} broadband filters centred at restframe 
1500\AA, 2200\AA, 3300\AA, and 4400\AA. 
These wavelengths are selected to correspond to the central wavelengths
of the filters used in our HST/ACS imaging survey of local starbursts 
(Hayes et al. \cite{ref:hayes05}).
In this study we found we could find non-degenerate fits in age --
\ebv\ space 
if we sampled the UV continuum slope and 4000\AA\ break. 
The same technique is adopted whereby we fit the synthetic SB99 spectra
to the `real' data.
A standard least-squares algorithm is used for different ages and internal 
reddenings using a given reddening law. 
From the best-fitting synthetic spectrum, the flux ratio between 
the $(1+z)\cdot1500$\AA\ continuum filter and the narrowband on-line filter 
is computed.
This scale-factor is then used to scale the observed continuum flux at
1500\AA\ to the continuum flux in the spectral region sampled by the 
on-line filter.  
\end{enumerate}

%
\section{The simulations\label{sect:sim}}

Our simulations aim to understand the way in which astrophysical 
conditions in the host-galaxy and inter-galactic medium (IGM) manifest 
themselves in the determination of \flya\ and \wlya.
To this end, we devised a number of different tests, varying
individually the parameters that describe the host galaxies, 
in order to examine their impact. 
In order to standardise the parameter setup, a `standard' restframe 
template was defined, the key parameters of which can be seen in 
Tab.  \ref{tab:stdpar}. 

\begin{table}[htbp]
\caption{The parameters of the `standard' restframe SED}             
\label{tab:stdpar}
\flushleft
\begin{tabular}{r l}        
\hline\hline                 
Parameter & Value \\ 
\hline                        
\wlya & 100\AA\ \\ 
Metallicity  & 0.001 \\
IMF $\alpha$ & -2.35 \\
IMF $M_\mathrm{low}$, $M_\mathrm{up}$  & 1.0$M_\odot$, 100$M_\odot$ \\
Age & 4Myr \\
Burst mode & Instantaneous \\
Composition & Stellar + Nebular\\ 
Secondary population & No \\
\hline                       
\end{tabular}
\flushleft
Note. IMF of the form: $dN/dM \propto M^{-\alpha} dM$ \\
\end{table}

By default, single stellar populations are treated, with the inclusion of
stellar and nebular emission, the relative contributions of  which are 
left unchanged. 
Unless otherwise stated, the equivalent width of the \lya\ line added 
to the spectrum is 100\AA\ for ease of visualisation (I.e.
fractional/percentage deviations are easily interpreted). 
No internal or Galactic extinction is applied to the spectra by default. 

We determined it necessary to run tests to study the impact of redshift
and \ion{H}{i} absorption along the line-of-sight, internal dust 
reddening, and mixing of stellar populations. 
Burst age is not considered in this paper for the following reason:
our preliminary tests showed that the recovery of \flya\ and
\wlya\ by all four techniques were self-consistent to within 2\%, (no age 
dependence on the recovered observables with age) over the first 200Myr. 
In addition, 
Charlot \& Fall (\cite{ref:cf93}) 
demonstrated \lya\ in emission can only be expected from considerably
younger star-forming galaxies of ages $\le 40$Myr; over this time we found 
evolution in the observables below the 1\% level.

Technique \#2 allows for the UV continuum slope (\bet) to be arbitrarily
chosen. 
In a large sample of LBGs re-sampled into quartiles according to \wlya, 
Shapley et al. (\cite{ref:shapley03})
find median values of $\beta=-1.09$ for the strongest \lya\ emitters
(upper quartile) with the continuum flattening to $\beta=-0.73$ for the 
weakest, lower quartile where the median \wlya\ is {-14.92\AA}. 
In contrast, the Starburst99 models show \bet\ in the range -2.6 to -2.0
over the first $\sim50$Myr. 
Excess flux in a narrowband filter $(n-b < X)$ is a typical selection 
function although the the amount of excess $X$ is arbitrarily chosen.
$X=0$ would correspond to a flat continuum, whereas $X<0$ would imply the
continuum level is rising towards \lya.
While $X$ can be computed by making some simple assumptions, the range 
in the power-law index (\bet) from which $X$ would be computed
can be very large. 
Without any additional data to constrain the continuum slope, any
assumption is ad hoc but, for the sake of simplicity, in the simulations 
presented here we assume $\beta=0$ (i.e. a flat continuum). 
Moreover, only a modest dust content is required to flatten the steep
continuum slopes observed in the FUV.
Additionally, it should be mentioned that the Equ. \ref{eq:fluxew} 
(technique \#1) only holds exactly if $\beta=0$, although the total 
spectral region sampled by this technique is much smaller.

\subsection{Cosmic distance and the IGM\label{subsect:IGM}}
Any two restframe SEDs generated from the same initial parameters
are identical.
The only way two identically created but redshifted SEDs can differ is
via the differing effects that result from intervening \ion{H}{i} in the IGM. 
The $z$-distribution and column density of \ion{H}{i} clouds is
pseudo-randomly generated, and as a result the redshifted SEDs can
differ greatly.
Thus a Monte-Carlo-type approach is adopted. 
A population of 10\,000 restframe SEDs is generated which, after
redshifting, differ only bluewards of \lya\ 
(except for the possible red wing of DLA systems that may extend 
on to the red side of the line -- the absorption line-centre is still
on the blue side).
For this population, mean and median averages of the fluxes and 
equivalent widths are then computed using the techniques described in
Sect. \ref{sect:fluxcomp}.
These computations are performed for populations at a range of redshifts
between zero and $z=9.0$, the highest redshift at which narrowband
survey techniques have been employed in an attempt to find LAEs
(Willis \& Courbin, \cite{ref:willis05}).
In a real imaging survey, there must be additional selection criteria 
by which a candidate is determined to be an emitter. 
In this case, we also recompute these mean values after the rejection
of all objects for which restframe \wlya\ value was determined to be 
less than 20\AA. 
This value is hereafter referred to as the ``refined mean".

It would also be interesting to simulate more realistic samples of 
galaxies. 
The \lya\ luminosity function has been compiled at $z=5.7$ and 6.5
by 
Malhotra \& Rhoads (\cite{ref:malrhoads04}). 
However, while spectroscopic observations were used in the compilation
of these luminosity functions for statistical purposes, the luminosities
of the LAEs themselves have been derived photometrically. 
Hence, they may well contain observational biases of the type we attempt
to address here.
Additionally, we would have to assume the characteristics of the 
observation itself such as sky-noise and 
it was deemed more revealing to study populations of identical objects. 

\subsection{Internal dust reddening}
In the absence of intervening \ion{H}{i} absorption, we study the
effect of internal dust extinction on the reliability of our recovery of
\flya\ and \wlya. 
Using the common starburst extinction laws 
[ Starburst: Calzetti et al. (\cite{ref:calzetti00});  
SMC: Pr\'evot et al. (\cite{ref:prevot84}); 
LMC: Fitzpatrick et al. (\cite{ref:fitz85}) ]
we artificially redden the template SEDs and investigate how reliably
the various techniques estimate \lya\ and \wlya. 
Technique \#4 relies upon fitting age and internal reddening in order
to estimate the continuum flux at \lya. 
In these experiments, we also investigate the reliability of this technique when
different extinction laws are used for the SED fitting from the one used to 
redden the intrinsic spectrum. 
This allows us to investigate how well \lya\ observables are recovered if an
inappropriate reddening law were assumed. 
For example, we redden the intrinsic SED with the SMC law and use the Calzetti
law in the SED fitting routines. 
We then investigate whether possible improvements can be made by 
including the reddening law itself as a free parameter in the fitting 
routine. 
Tests are carried out in the range $0.0 < E_{B-V} < 0.5$ at redshifts of
2, 4, and 6.

\subsection{Underlying stellar populations\label{sect:msp}}

Again without the effects of intervening \ion{H}{i} systems, we 
investigate how accurately \flya\ and \wlya\ are recovered when the
\lya\ emitting object hosts varying components from an underlying
population. 
This is mainly relevant to technique \#4 which relies upon sampling of
the 4000\AA\ break, and, while we don't study \flya\ or \wlya\ {\em vs.}
age directly, this study addresses the effect an aged stellar population
may have.
As before we adopt the default template spectrum as a `primary' 
SED 
(I.e. the single stellar population starburst spectrum defined in Tab.
 \ref{tab:stdpar} with an age of 4Myr).
We then add an assortment of older, post-starburst spectra to the
primary, scaling to varying normalisation factors at a
wavelength of 4500\AA\ (i.e. scaling by the approximate $B-$band 
luminosity).
We thereby artificially create young starburst galaxy spectra with a
population of older stars. 
This normalisation factor is designated $n_{4500}$ and, by this
definition: $n_{4500}=0$ is just the default SED defined in Tab.
\ref{tab:stdpar}; $n_{4500}=1$ means the default and aged populations
contribute equally at 4500\AA; and $n_{4500}=10$ means the old
population is a factor of 10 more luminous than the starburst SED at
4500\AA. 
The parameters used in the generation of SED of the underlying
population can be seen in Tab. \ref{tab:mspset} but briefly, we perform
tests to examine the age of the underlying population, its metallicity,
and its normalisation coefficient, $n_{4500}$.
Then, after redshifting the SEDs, we investigate what values of \flya\
and \wlya\ our techniques recover. 
When the SED-fitting technique is used (technique \#4), we always used
the unmodified set of starburst templates for fitting. 
Tests are performed at redshifts of 2, 4, and 6. 
Tab. \ref{tab:mspset} shows the different old stellar populations
and their contributions as applied to the standard setup, along
with the resulting \wlya\ calculations.



%
\section{Results and discussion\label{sect:resdis}}

\subsection{Cosmic distance and IGM\label{sect:resdis:IGM}}

\begin{figure*}[htbp]
\centering
\includegraphics[width=8.0cm,height=4.0cm]{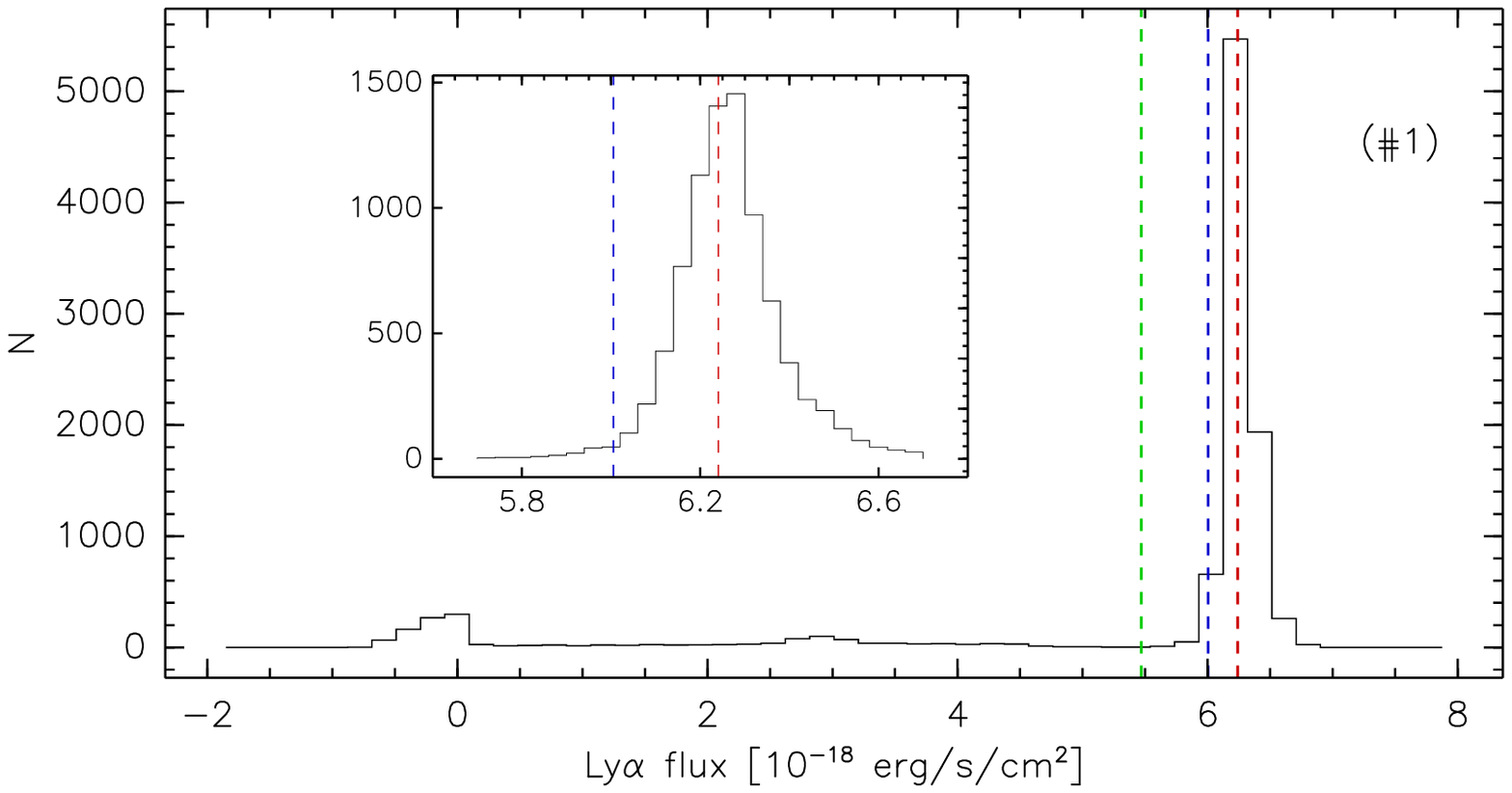}
\hspace{0.3cm}
\includegraphics[width=8.0cm,height=4.0cm]{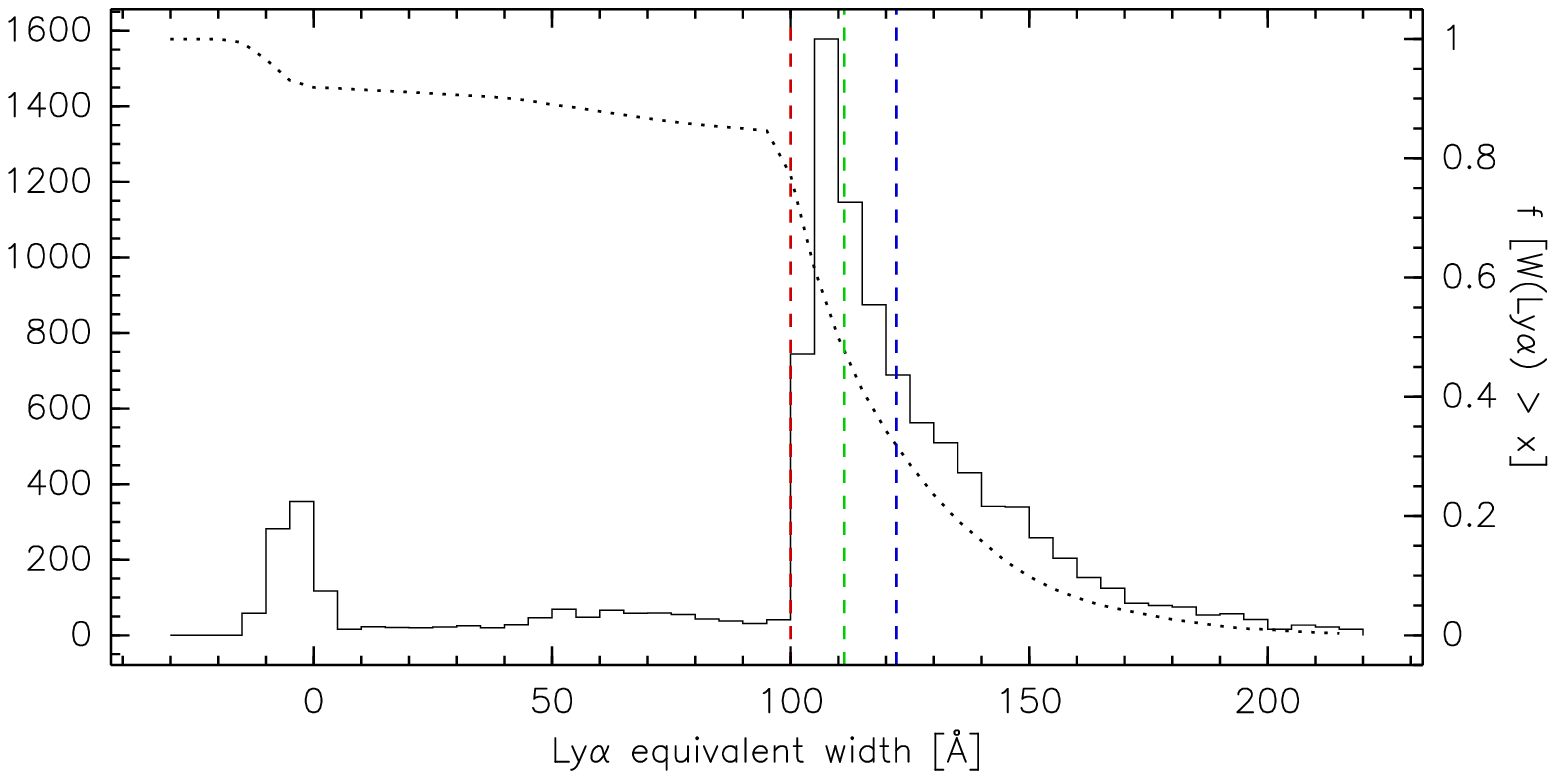}

\vspace{0.3cm}

\includegraphics[width=8.0cm,height=4.0cm]{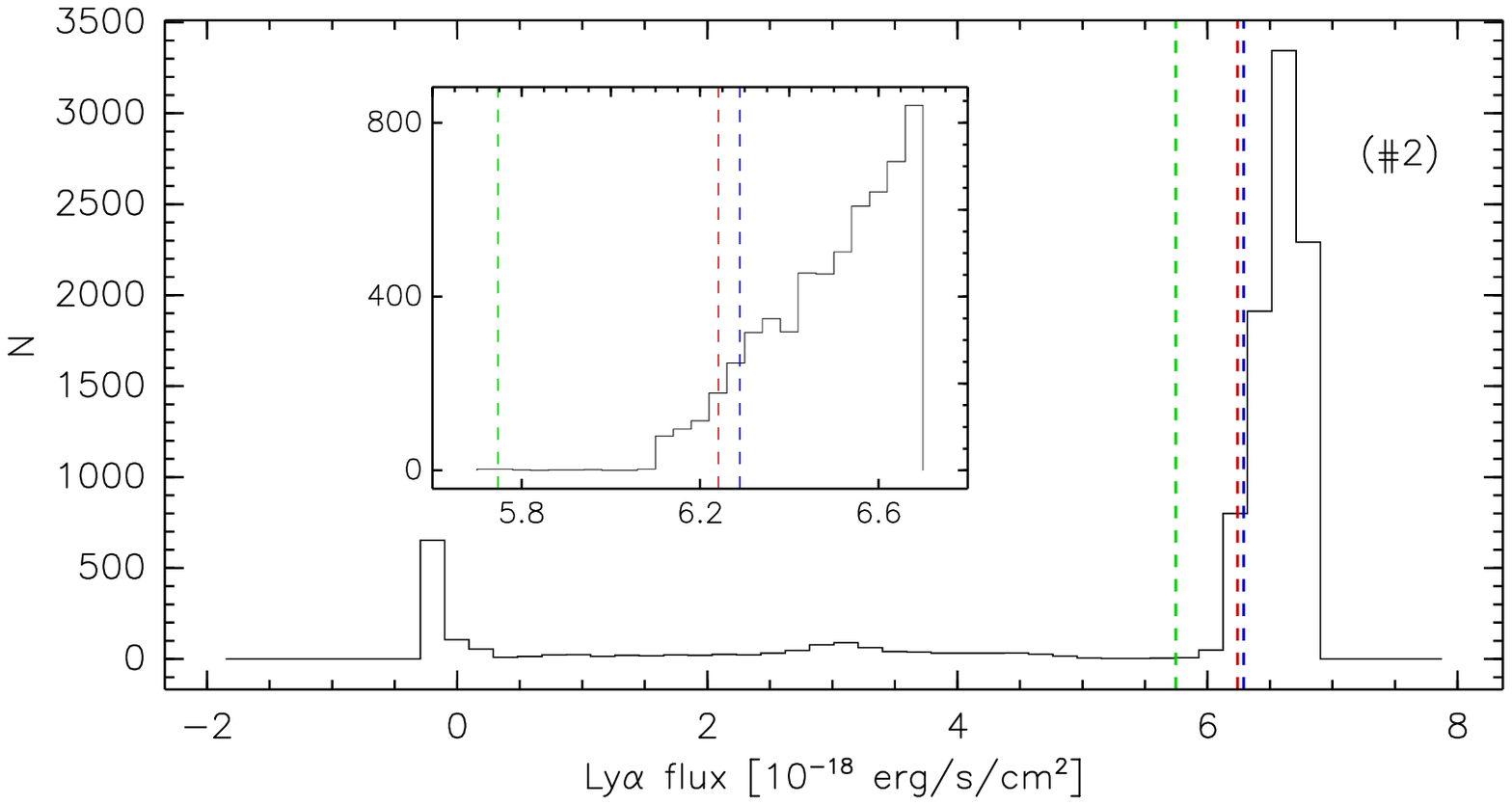}
\hspace{0.3cm}
\includegraphics[width=8.0cm,height=4.0cm]{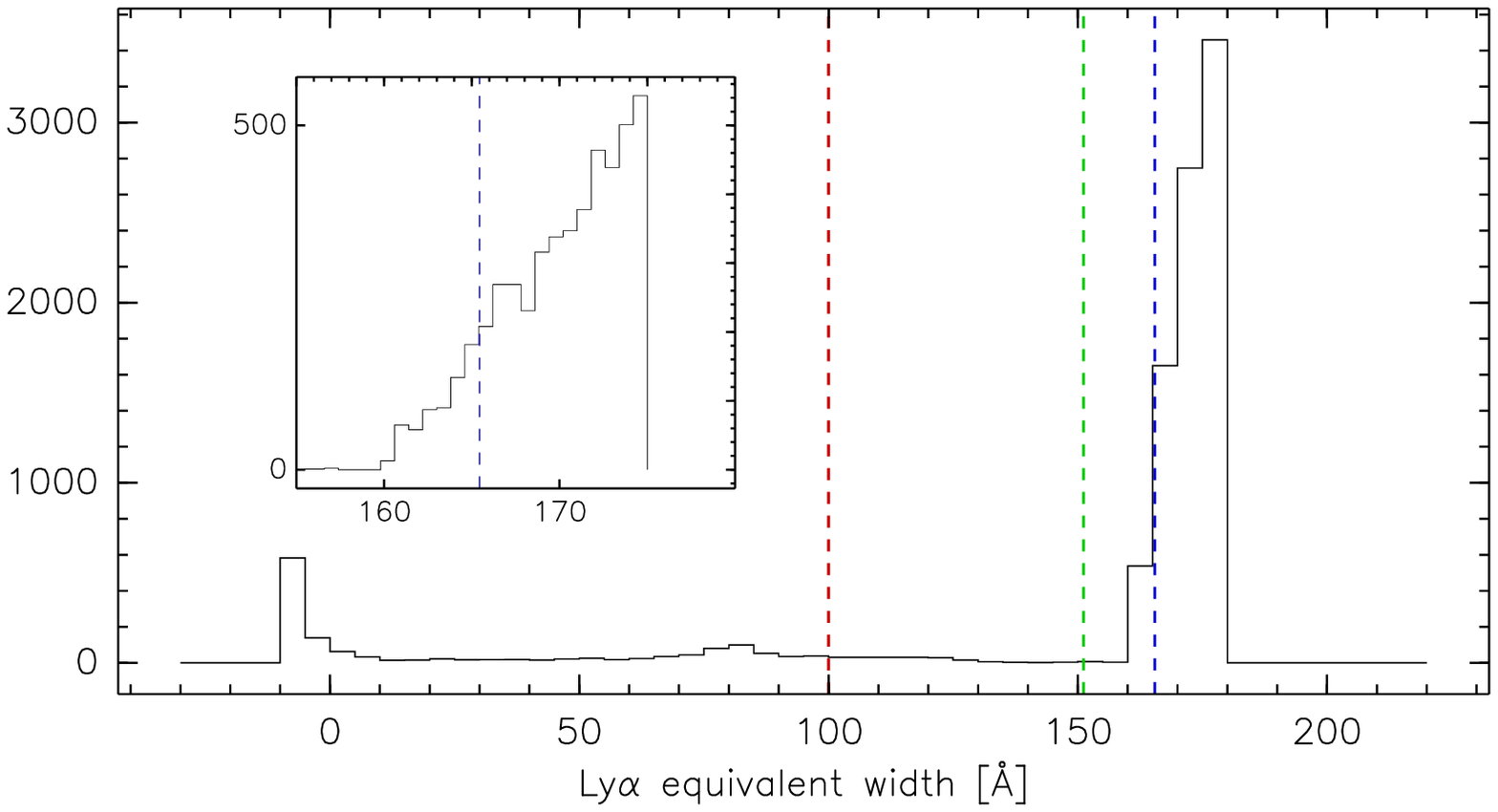}

\vspace{0.3cm}

\includegraphics[width=8.0cm,height=4.0cm]{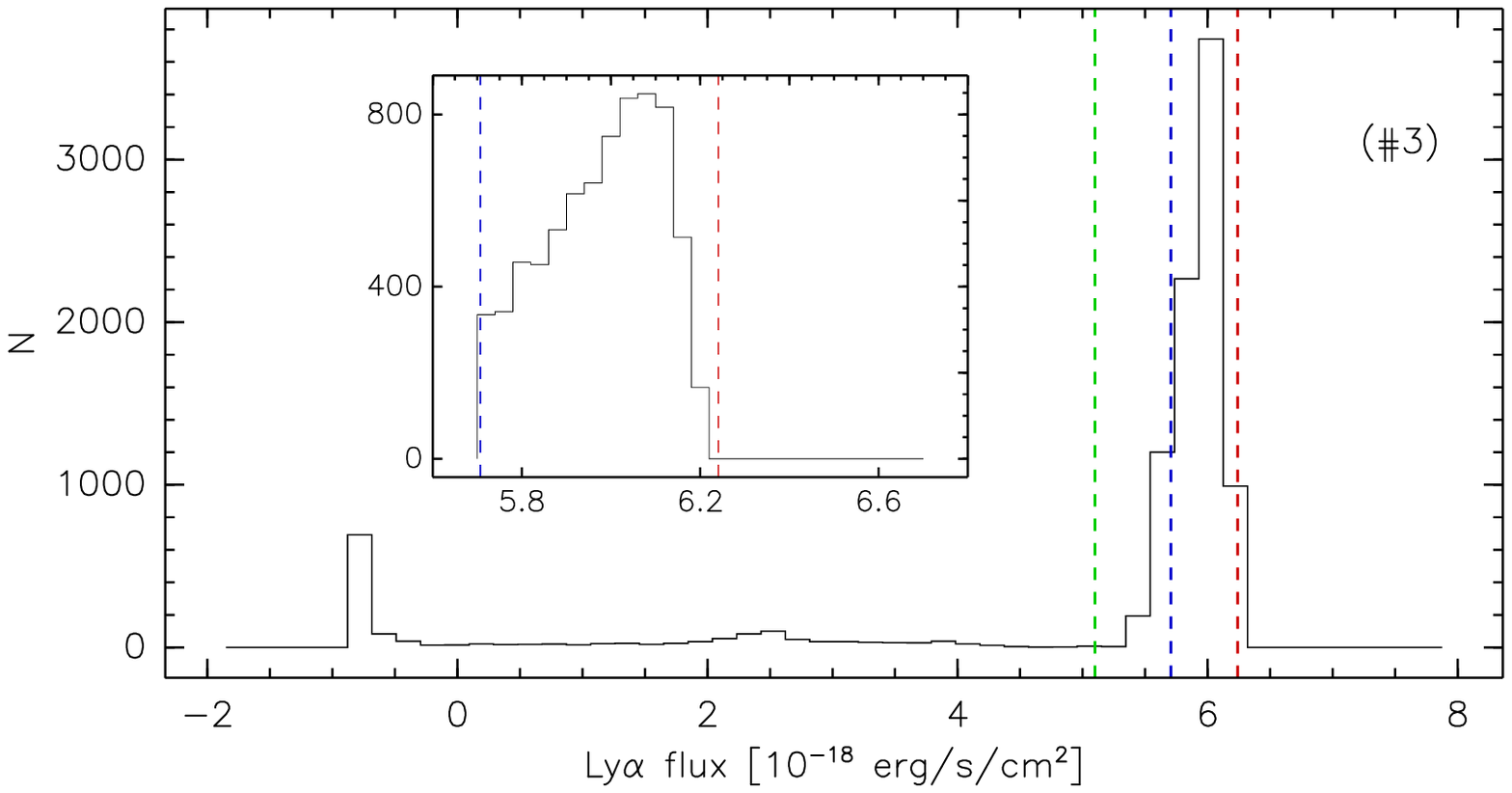}
\hspace{0.3cm}
\includegraphics[width=8.0cm,height=4.0cm]{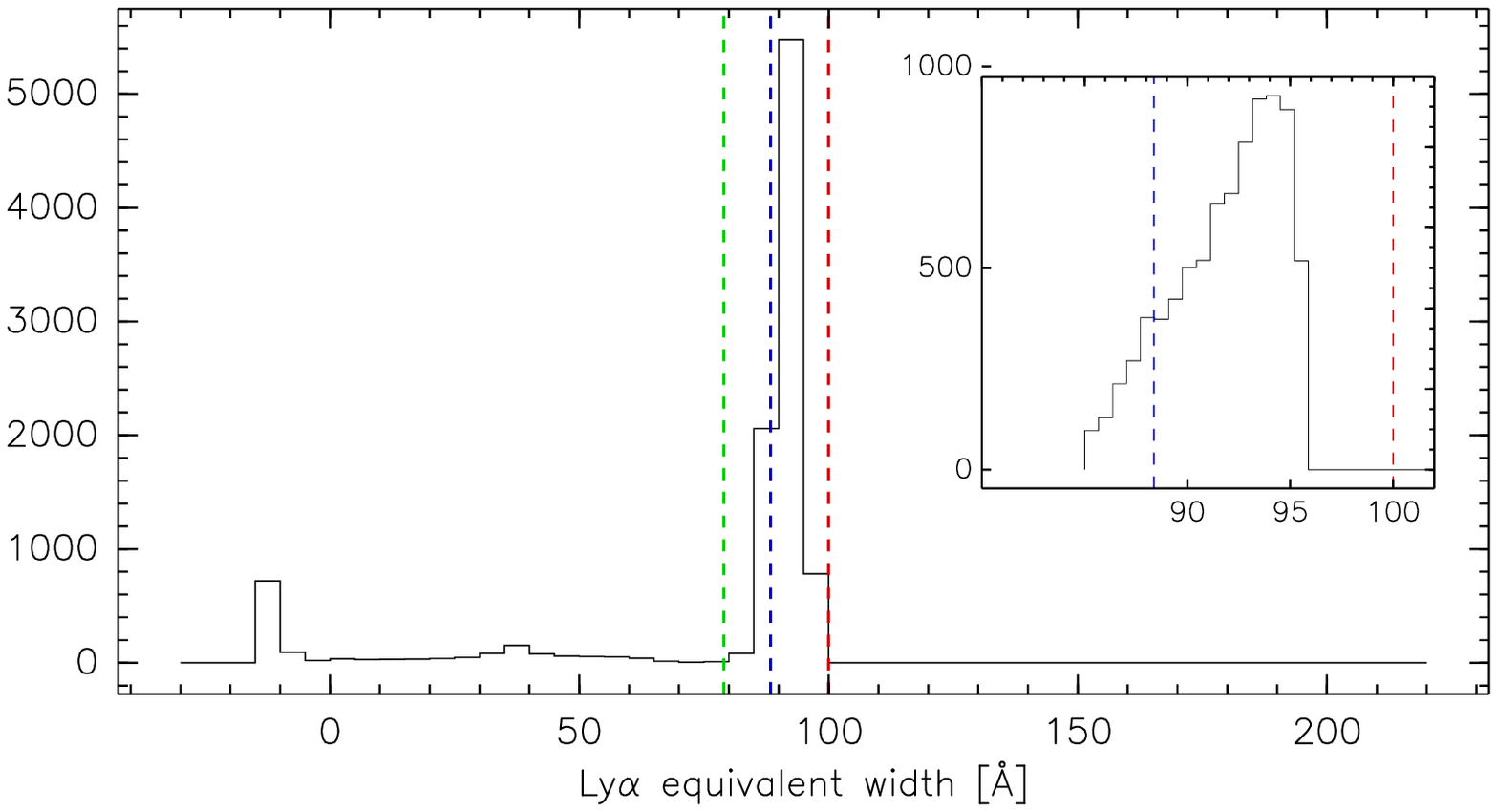}

\vspace{0.3cm}

\includegraphics[width=8.0cm,height=4.0cm]{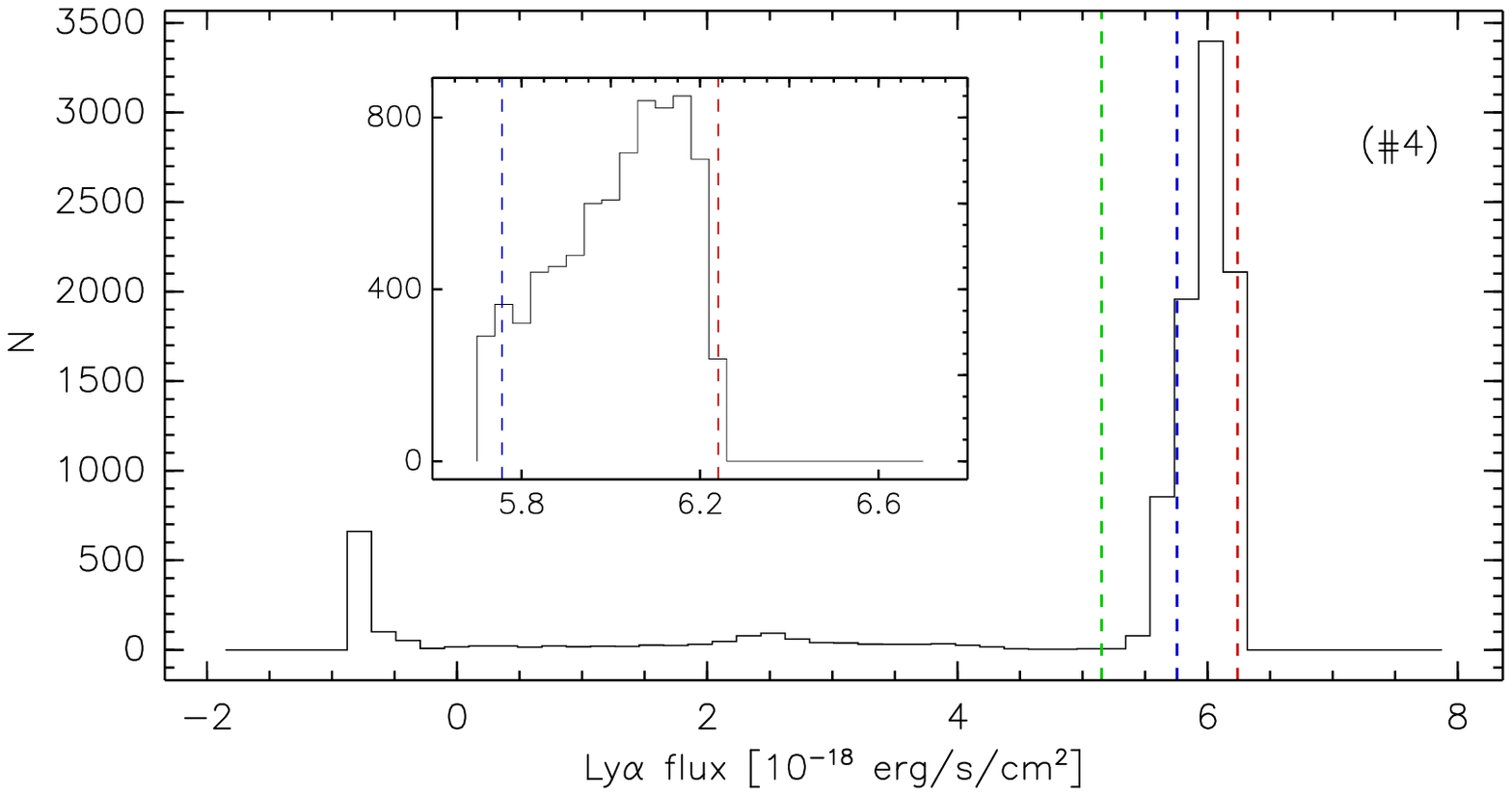}
\hspace{0.3cm}
\includegraphics[width=8.0cm,height=4.0cm]{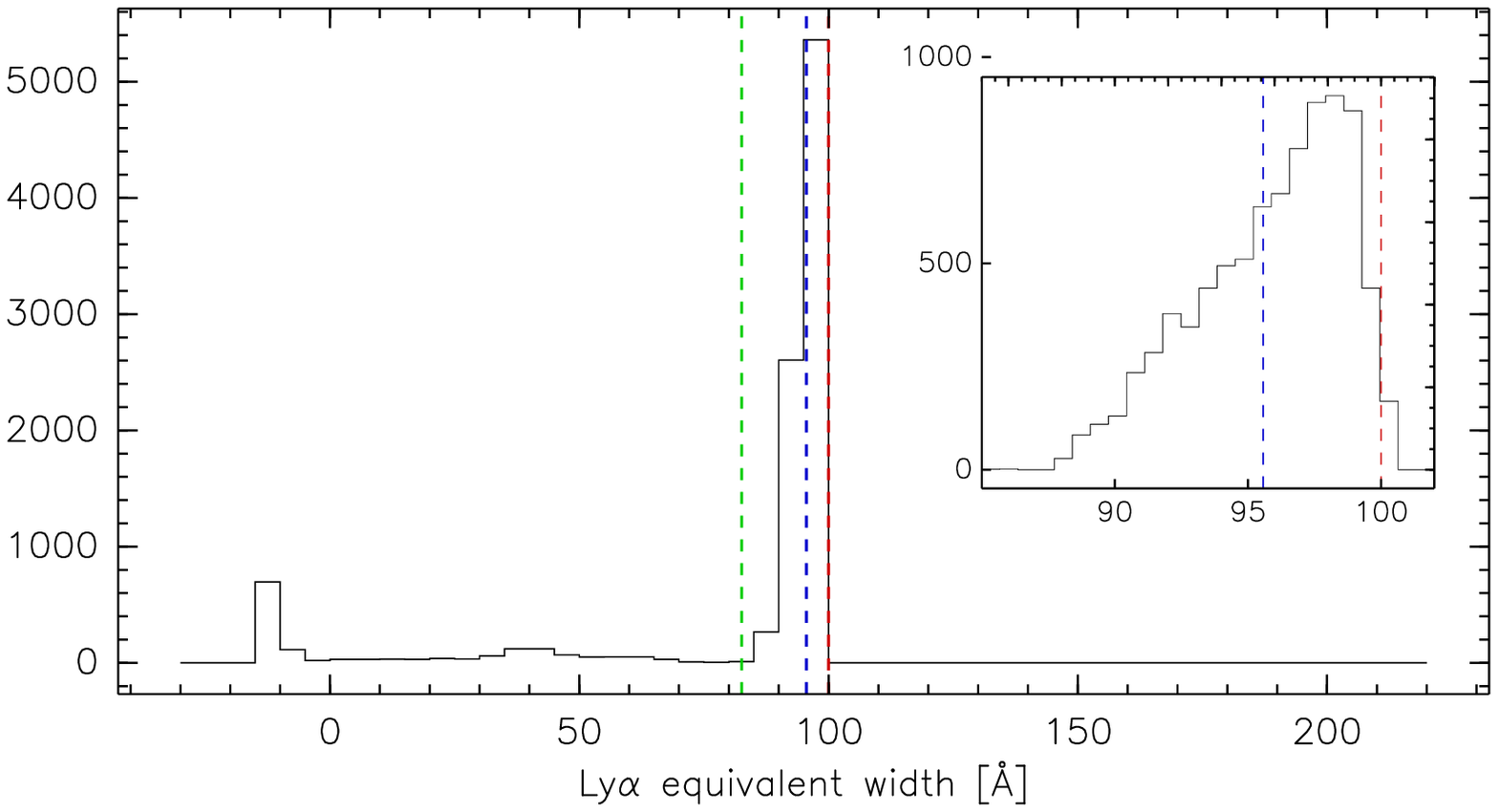}

\caption{Histograms showing the distribution of detected \lya\
flux (left) and equivalent width (right) for a population of 10\,000
galaxies at a redshift of $z=5.7$. 
The descending pairs of plots represent distributions as computed by
observational techniques \#1, \#2, \#3, \#4. 
The vertical dashed red lines indicate the true line flux and
equivalent width (I.e. that added to the restframe SED).
The vertical green line shows the mean values computed for the whole
sample of 10\,000 objects, while the blue line shows the mean value of
these quantities for all objects with \wlya$>20$\AA\ (refined mean). 
The dotted black line in the uppermost equivalent width plot shows the 
fraction of galaxies in the sample with observed \wlya\ greater than the 
value on the abscissa axis (I.e. for that value of \wlya, the fraction
of galaxies that have \wlya\ greater than this value). 
[{\em See the electronic journal article for the colour version of this figure.}]
}
\label{fig:hist_z5.7}
\end{figure*}

As described in Sect. \ref{sect:fluxcomp} and Sect. \ref{subsect:IGM} we 
generate populations of galaxies at a given redshift, and
determine the mean, median, and refined mean values of observed 
\flya\ and \wlya\ for this population using the four observational 
techniques. 
Fig. \ref{fig:hist_z5.7} shows a set of histograms representing the 
distribution of these observed quantities for a population of 10\,000
galaxies 
at $z=5.7$, selected to correspond to the redshift of number of \lya\
surveys (eg. 
Thommes et al. \cite{ref:thommes98}; 
Rhoads \& Malhotra \cite{ref:rhoads01}; 
Westra et al. \cite{ref:westra05}).
Restframe \wlya\ is 100\AA. 

A noteworthy feature, present in all these plots is the population of
$\sim 750$ 
galaxies with \flya\ and \wlya\ around zero (and slightly
negative). 
In the line-of-sight to these objects a DLA system has been generated
near to the target LAE, the red damping wing of which has removed the
\lya\ line from the spectrum. 
The slight negativity of these values comes from the fact that, in order
to estimate the line-only flux, the continuum has been subtracted and,
in all these cases, continuum subtraction has resulted in negative
fluxes in the line.
These objects would not be found by imaging observations that target
emitters and hence
including them in the computation of average quantities would be
misleading.
This effect therefore leads to a redshift-dependent
incompleteness, which at $z=6$ is about 10\% for these filters. 
Such effects would need to be treated in the computation of the \lya\ luminosity
function. 
It is for this reason that we compute refined mean values for the 
population of objects with detected \wlya$>20$\AA\ only. 

Dashed vertical lines have been added to all these histograms in three
colours. 
The red line shows the flux or equivalent width of the \lya\ line that was
added to the restframe spectrum, the green line shows the mean of the 
recovered quantities, and the blue line shows the refined mean.

The top pair of plots in Fig.  \ref{fig:hist_z5.7} represent the
quantities as determined using technique \#1 (a narrow and broad filter
centred at $(1+z) \cdot \lambda_{\mathrm{Ly}\alpha}$). 
The \flya\ histogram shows a strong peak in the distribution that 
perfectly corresponds to intrinsic line flux. 
The mean of the distribution falls 15\% short of this value.
The coincidence of the peak in flux distribution and intrinsic flux
demonstrates the power of this method in determining line flux. 
The top right panel in this figure shows the distribution of \wlya\ when
computed by this technique (\#1). 
While \flya\ is accurately recovered and evenly distributed around the
correct value, the \wlya\ distribution cuts on
sharply at \wlya=100\AA, peaks, and exhibits a long tail, extending to
220\AA. 
While some flux may be removed from the narrowband filter by intervening
\ion{H}{i} clouds, the broadband filter samples the continuum
significantly bluewards of \lya.
Hence \ion{H}{i} clouds in the IGM can remove a significant amount of flux 
from this broadband observation. 
Consequently, the continuum flux at \lya\ can be drastically underestimated,
resulting in a doubling of observed \wlya\ in extreme cases. 
The black dotted line in this plot shows the fraction of objects with
recovered \wlya\ greater than the value on the abscissa. 
It crosses the red dashed line (100\AA) at 0.8, implying that \wlya\ has
been overestimated for 80\% of {\em all} the objects (including those
with \wlya$\sim 0$ that represent the incompleteness in the observed
population).
After removing the \wlya$\sim 0$\AA\ objects, this fraction is more like
90\%. 

The second pair of plots represents the distribution of the same detection 
properties, as determined by technique \#2. 
Here the same on-line filter is used but a single broadband filter 
centred at $(1+z)\cdot1500$\AA\ is used to estimate the
continuum at \lya\ by assuming $\beta=0$ between the spectral regions
sampled by these filters.
Since our standard restframe template is that of a 4Myr starburst with
no dust, the continuum is increasing towards the line
($\beta \sim -2.6$)
and the subtraction of an underestimated continuum results in a slight 
shift of the distribution towards higher fluxes.  
In this case where the line is strong, the measured flux in the
narrowband filter is dominated by the line, and the relative effect of
underestimating the continuum is small. 
We have verified that the effect is more significant when smaller rest 
equivalent widths are input (i.e. continuum-dominated observations). 
In this example, underestimating the continuum 
flux has a more pronounced impact upon the estimate of \wlya\ and  
the mode in the \wlya\ distribution is shifted from 100\AA\ to
$\sim 170$\AA.
While technique \#2 may provide reliable line-fluxes when the line is 
strong, any equivalent widths derived in this manner cannot be
considered robust. 
Even if there is no line (\wlya$=0$), for a young, unreddened burst 
with $\beta < 0$, the assumption of $\beta = 0$ will result in a positive 
observed equivalent width and false positive detection of narrowband excess.

Techniques \#3 and \#4 reliably reproduce both \flya\
and \wlya\ in this study. 
Technique \#3 assumes the UV continuum slope to be a power-law between
2200\AA\ and \lya\ and uses off-line filters placed at restframe 1500 
and 2200\AA\ to extrapolate to \lya. 
Technique \#4 makes no assumptions about the continuum, and uses a
4-filter SED-fitting technique to estimate the continuum level at \lya. 
With the exception of the small peaks around \flya\ and \wlya=0, these
distributions are strongly peaked around their intrinsic values. 
The narrow distribution of \wlya\ 
demonstrates how successful these techniques are at estimating the
continuum fluxes at line-centre.
The refined mean values of \wlya\ are in error by only $\sim5$\%. 

\begin{figure*}[htbp]
\centering
\includegraphics[width=8.0cm,height=4.0cm]{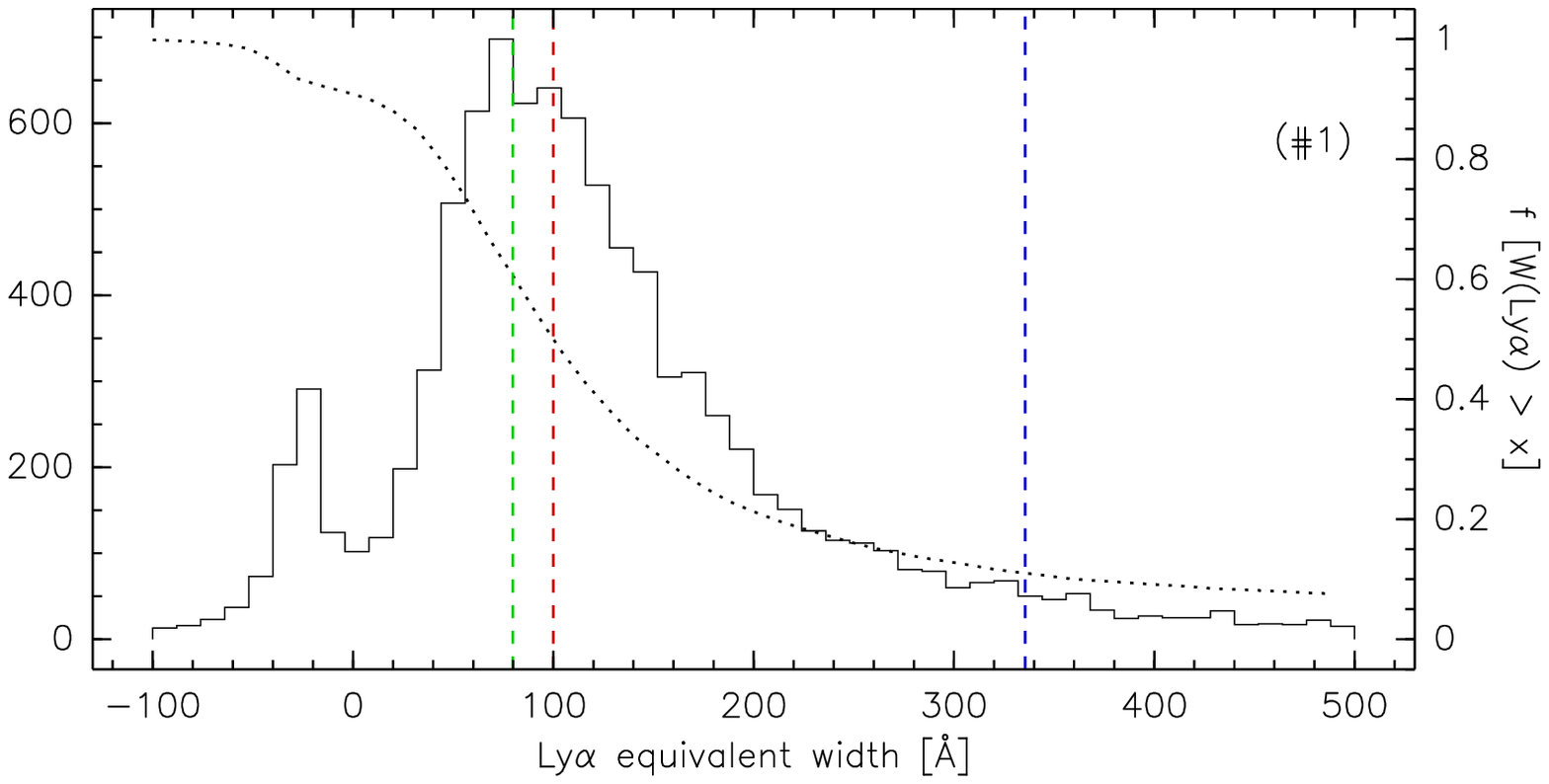}
\hspace{0.3cm}
\includegraphics[width=8.0cm,height=4.0cm]{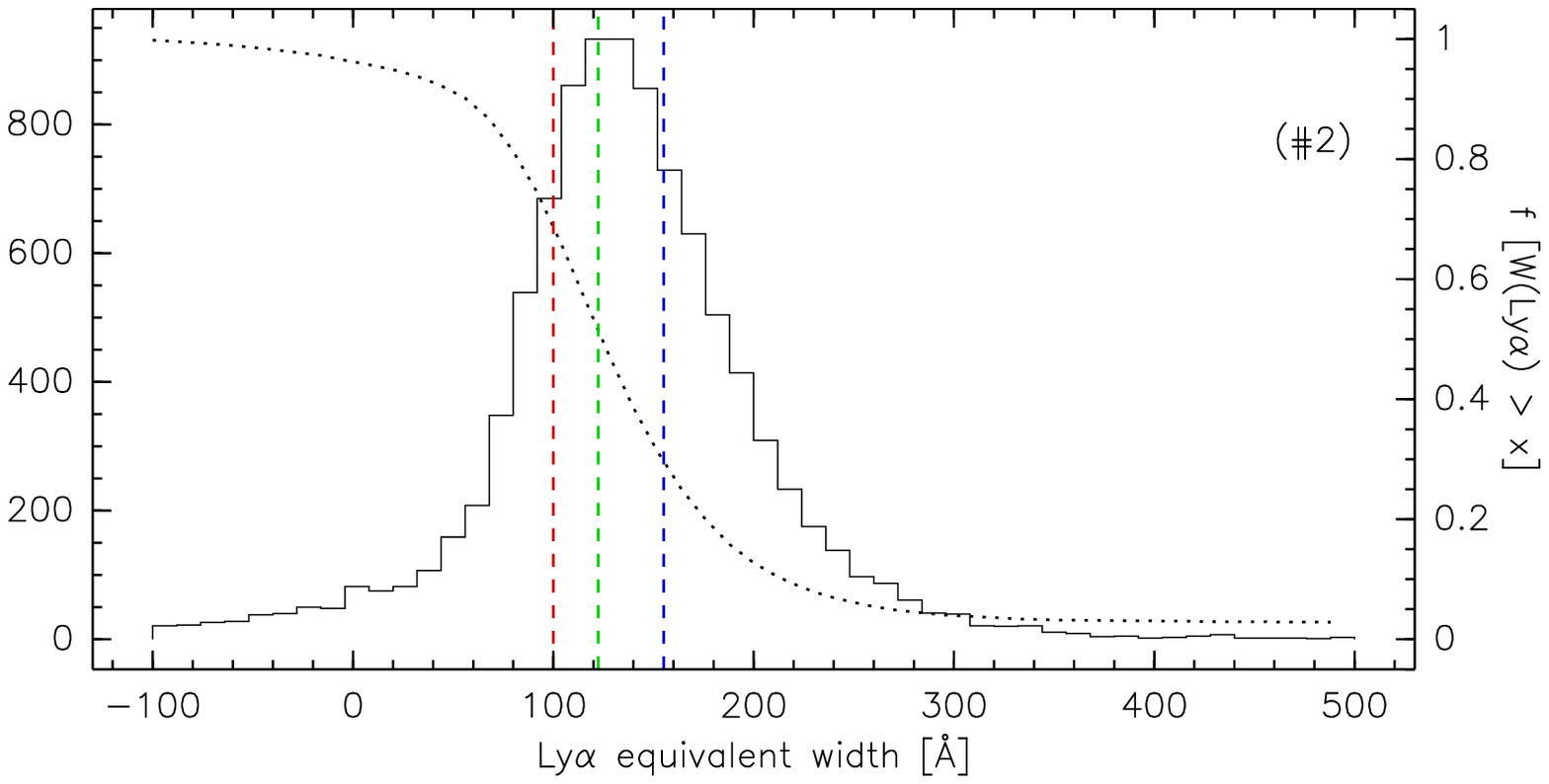}

\vspace{0.3cm}

\includegraphics[width=8.0cm,height=4.0cm]{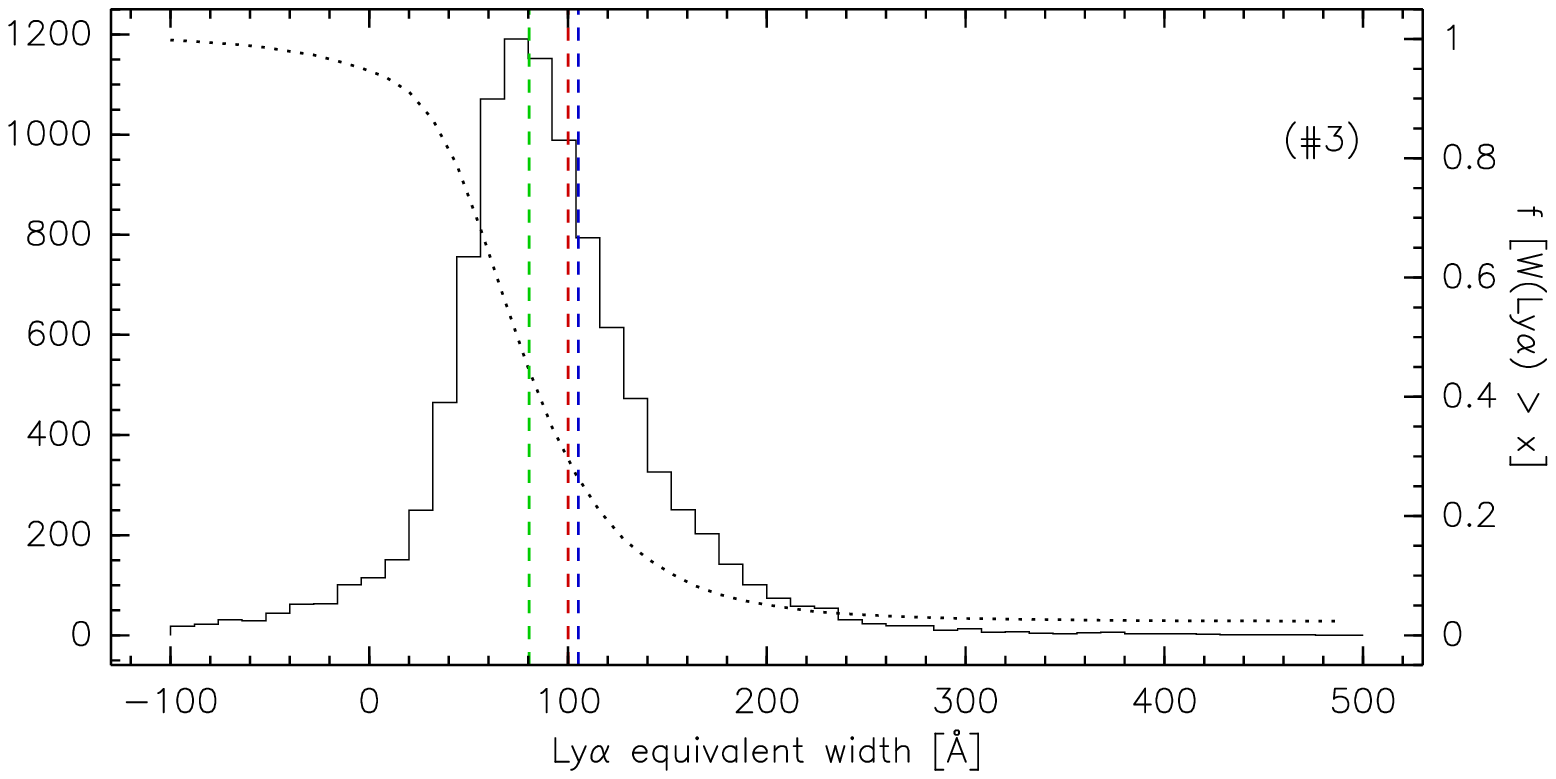}
\hspace{0.3cm}
\includegraphics[width=8.0cm,height=4.0cm]{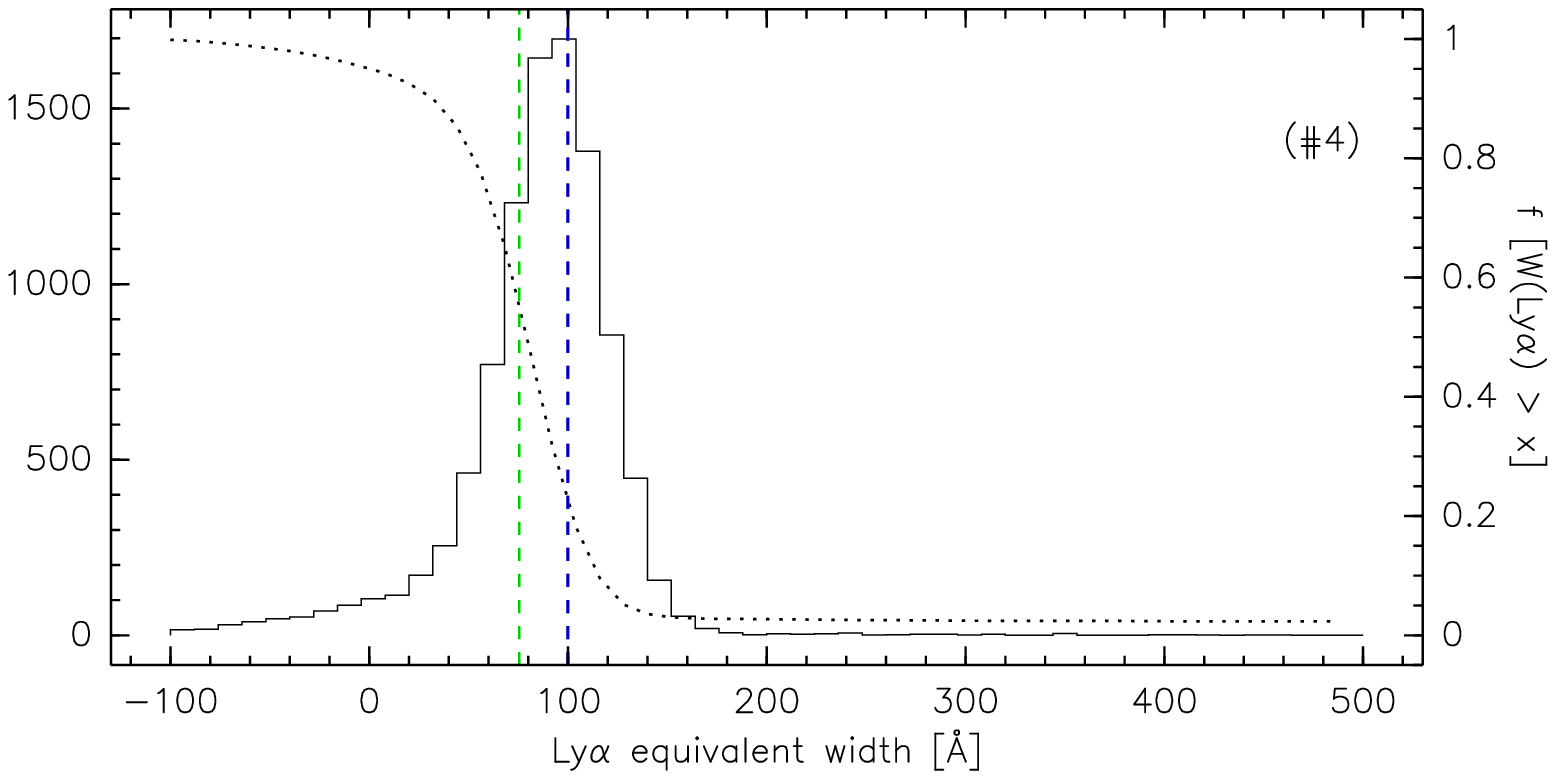}
\caption{Histogram showing the detected \wlya\ of 10\,000 galaxies at
$z=5.7$. 
These are the same objects as plotted in Fig. \ref{fig:hist_z5.7} but 
fluxes have been regenerated within assumed 
observational errors from a simple assumed noise model. 
Dotted line and vertical dashed lines as in Fig. \ref{fig:hist_z5.7}.
[{\em See the electronic journal article for the colour version of this figure.}]
}
\label{fig:hist_obserr}
\end{figure*}

In reality, all the accuracy of all these results would be dependent on
the recovered flux in the various bands and the sky-noise. 
While this is not strictly the angle we have chosen for this article, we
investigate and discuss the effects of sky noise below.
Fig. \ref{fig:hist_obserr} demonstrates how the various technique fare near 
the detection limit with the the application of a simple sky-noise model.

\begin{figure*}[htbp]
\centering
\includegraphics[width=8cm]{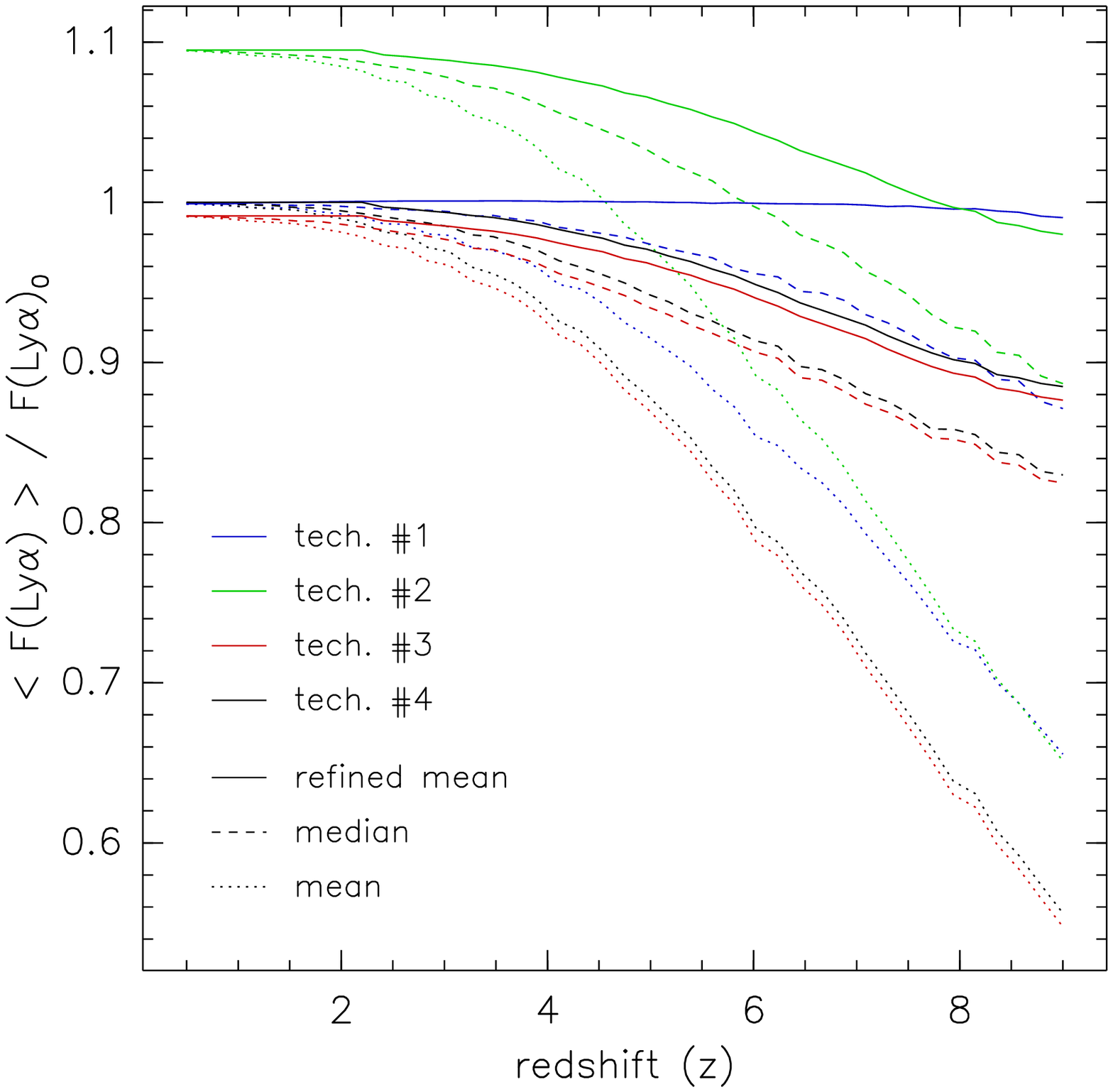}
\hspace{0.8cm}
\includegraphics[width=8cm]{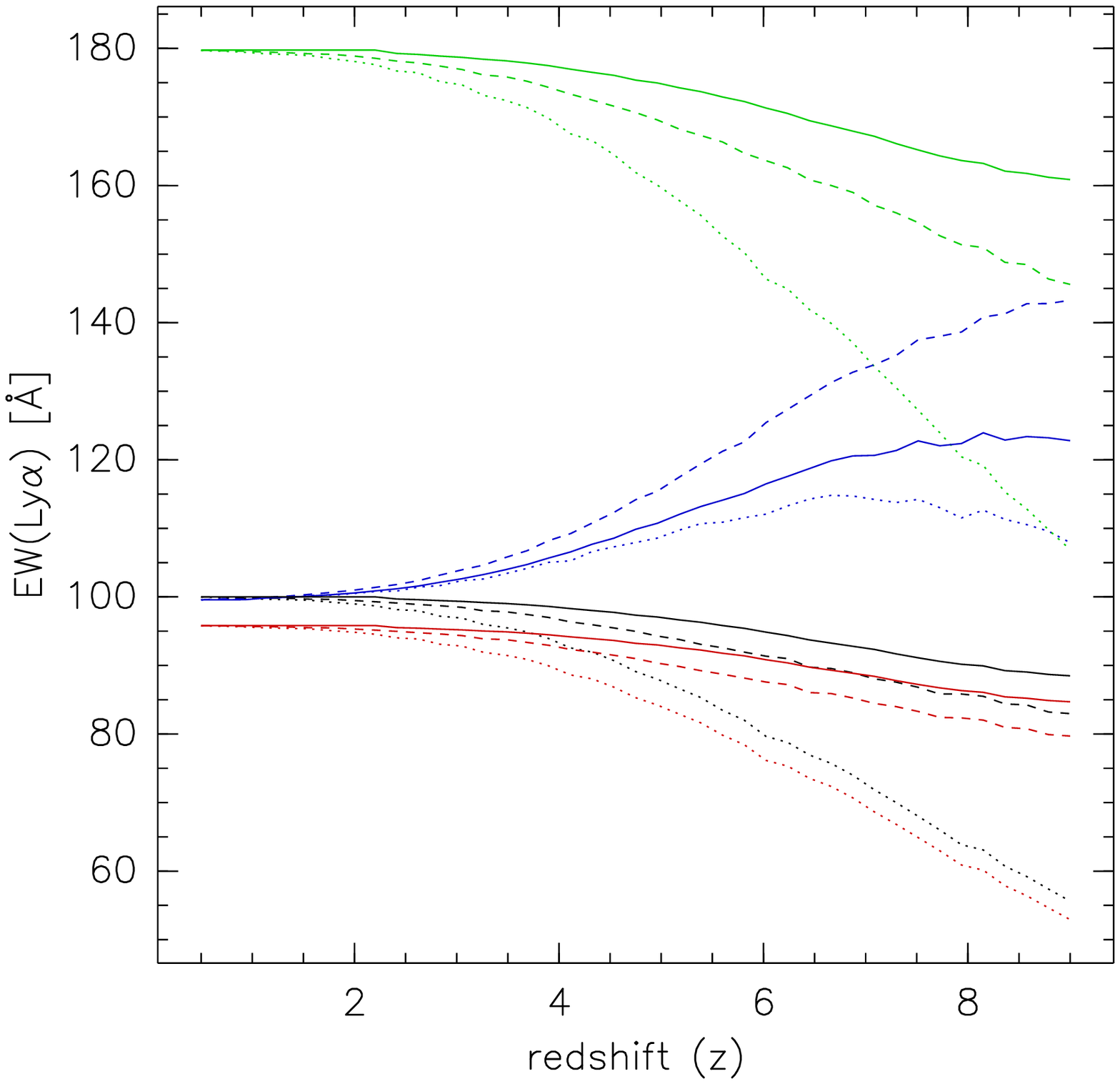}
\caption{The evolution of \flya\ and \wlya\ as function of redshift. 
The left plot shows the recovered \flya, normalised by the predicted flux
at the observatory, calculated from the intrinsic line luminosity 
added in
the restframe (I.e. the fractional deviation from the true value). 
The right plot shows the evolution of \wlya.
Colour blue represents technique \#1, green: technique \#2, red: technique
\#3 and black: technique \#4. 
Dotted lines represent total mean values, dashed lines represent median
values, and solid lines show the refined means. 
[{\em See the electronic journal article for the colour version of this figure.}]
}
\label{fig:w_vs_lafwithz}
\end{figure*}

Fig. \ref{fig:w_vs_lafwithz} shows the evolution of the total mean, refined
mean, and median values of \flya\ and \wlya\ as a function of redshift. 
The blue curves represent technique \#1, green curves technique \#2,
red curves technique \#3, and black curves technique \#4. 
The flux values are the average observed fluxes, normalised by the computed
intrinsic line
flux at the observatory (i.e. the intrinsic line-luminosity scaled down 
by the luminosity distance). 

As previously discussed, technique \#1 reliably recovers \flya\ in the
line-dominated limit and this is here demonstrated to be independent of 
redshift. 
However, \wlya\ as computed using this technique becomes progressively
worse with increasing redshift as can be seen from the right plot
-- the refined mean value for \wlya\ at $z=6.5$ is overestimated by 
20\%. 
This is a result of the increasing redshift-density of intervening 
clouds with redshift $(\propto (1+z)^{2.45})$, causing a systematic
decrease of the estimate of the continuum flux.
Technique \#2 overestimates \flya\ by $\sim10$\% at $z=2$. 
This results from the assumption of a flat spectrum not accounting for
the blue continuum slope of a young stellar population. 
This technique is unreliable in estimating \wlya\ and at all redshifts 
as the right plot shows, never comes close to estimating the correct \wlya.
Again as demonstrated in Fig. \ref{fig:hist_z5.7}, techniques \#3 and
\#4 both reliably determine \flya\ and \wlya\ at low and high redshift
-- the refined mean values of \wlya\ deviate from the expected values by
less than 9, and 5\%, respectively, at $z=6$.

It is, of course, possible to fit a function to any of the
redshift-evolution curves shown in Fig. \ref{fig:w_vs_lafwithz},
thus obtaining a functional prescription for the fractional under- or 
over-estimate of \flya\ and \wlya\ as a function of redshift. 
Such a formula would be highly desirable as it would be directly 
applicable to the results obtained by previous surveys. 
However such a corrective formula would not only be a function of
redshift, but also of the filter widths, response profiles, and 
positioning in wavelength. 
Of course, such information is well-known and we recommend simulations 
to be carried out on a survey-by-survey basis, and with appropriate
consideration  of the errors and selection criteria.

Malhotra \& Rhoads (\cite{ref:malrhoads02}) used a technique similar to 
technique \#1 to observe LAEs at a redshift of 4.5, uncovering a
population of galaxies with curiously high equivalent width (median 
\wlya$\sim 400$\AA ). 
Since there is no AGN activity associated with these objects (Wang et
al. \cite{ref:wang04}) and
the maximum \wlya\ available from star-formation (with `normal' IMFs and
metallicities) is 240\AA\  
 (Charlot \& Fall \cite{ref:cf93}),
such a discovery attracts speculation.
Using our simulations we determine that at $z=4.5$, the median value of 
\wlya\ may be overestimated by a factor of only $\sim12$\% -- clearly
Malhotra \& Rhoads'
high-\wlya\ result is not
an observational effect of the type under consideration in this article.
However, the effect of intervening \ion{H}{i} on \wlya\ as computed
by technique \#1 is that a large spread in \wlya\ is produced (see Fig.
\ref{fig:hist_z5.7}, top right), leading to 
a significant fraction of LAEs with overestimated \wlya\ -- 20\% of all the
objects have \wlya\ overestimated by more than 40\%. 
Certain selection functions (eg. narrowband excess) will then be more 
inclined to pick up these objects. 

In order to further investigate some more realistic observational effects, 
we implemented a simple noise model. 
Taking the fluxes obtained from the population of objects shown in Fig. 
\ref{fig:hist_z5.7} (10\,000 objects at $z=5.7$ with \wlya=100\AA), 
we assumed that the criterion for a `detection' is $5\sigma$. 
We then assigned a $S/N = 5$ to each narrowband flux, and weighted this by 
the ratio of detected \flya\ to the intrinsic line flux 
($F_{\mathrm{Ly\alpha,0}}$).
$S/N$ was assigned to the broadband observations in an identical manner:
5 weighted by the ratio of the observed flux to the intrinsic flux.
Note this weighting modification of $S/N=5$ is only applicable to the 
broadband filter centred at \lya\ -- the redder filters are not affected 
by \ion{H}{i} clouds in the IGM, and all have $S/N=5$. 
With $\sigma$ for all the observations, we used Box-Muller transforms to 
generate Gaussian deviates around the fluxes in each passband. 
We fed these fluxes back into the expressions used to derive \wlya\ 
and replotted the resulting distributions of \wlya\ which can be seen in 
Fig. \ref{fig:hist_obserr}.

In all cases, the main distribution is now significantly broadened;
the mode has shifted towards lower values of \wlya, and a high \wlya\
tail has been produced. 
This results from the fact that equivalent width is the quotient of two
values $(F_\mathrm{Ly\alpha} / f_{cont})$. 
Symmetrical redistribution of the denominator results in asymmetric 
redistribution of the combined quotient; compressed on the low side of the mean 
(lowering the mode) and extended at high values. 
This is most striking using technique \#1 since signal in the broad
filter is lost to \ion{H}{i} absorption in the IGM -- $\sim 20$\% of
objects have \wlya\ overestimated by a factor of 2. 
Still the number of objects scattered to very high \wlya\ is small 
overall, we reiterate that some selection criteria will include them: 
in a narrowband survey they are likely to exhibit clear narrowband excess.
Redistribution of \lya\ has caused the plots for all techniques to
appear similar: modes have been reduced slightly and extended, 
high-\wlya\ tails are present.
One noteworthy feature about the progression from technique \#1 through 
\#4 is the movement of the refined mean of the distribution (blue
vertical line) towards the intrinsic value of 100\AA, and for the same
reason, the steepening of the black dotted line. 
The addition of more filters is necessary to prevent the extreme spreading 
of the distribution and scattering to extreme values of \wlya.
For technique \#4 the refined mean value now accurately recovers the intrinsic
values of \wlya, and \wlya\ is overestimated by more than 50\% for very few
objects.

\subsection{Internal dust}

\begin{figure*}
\includegraphics[width=6.2cm,height=4.5cm]{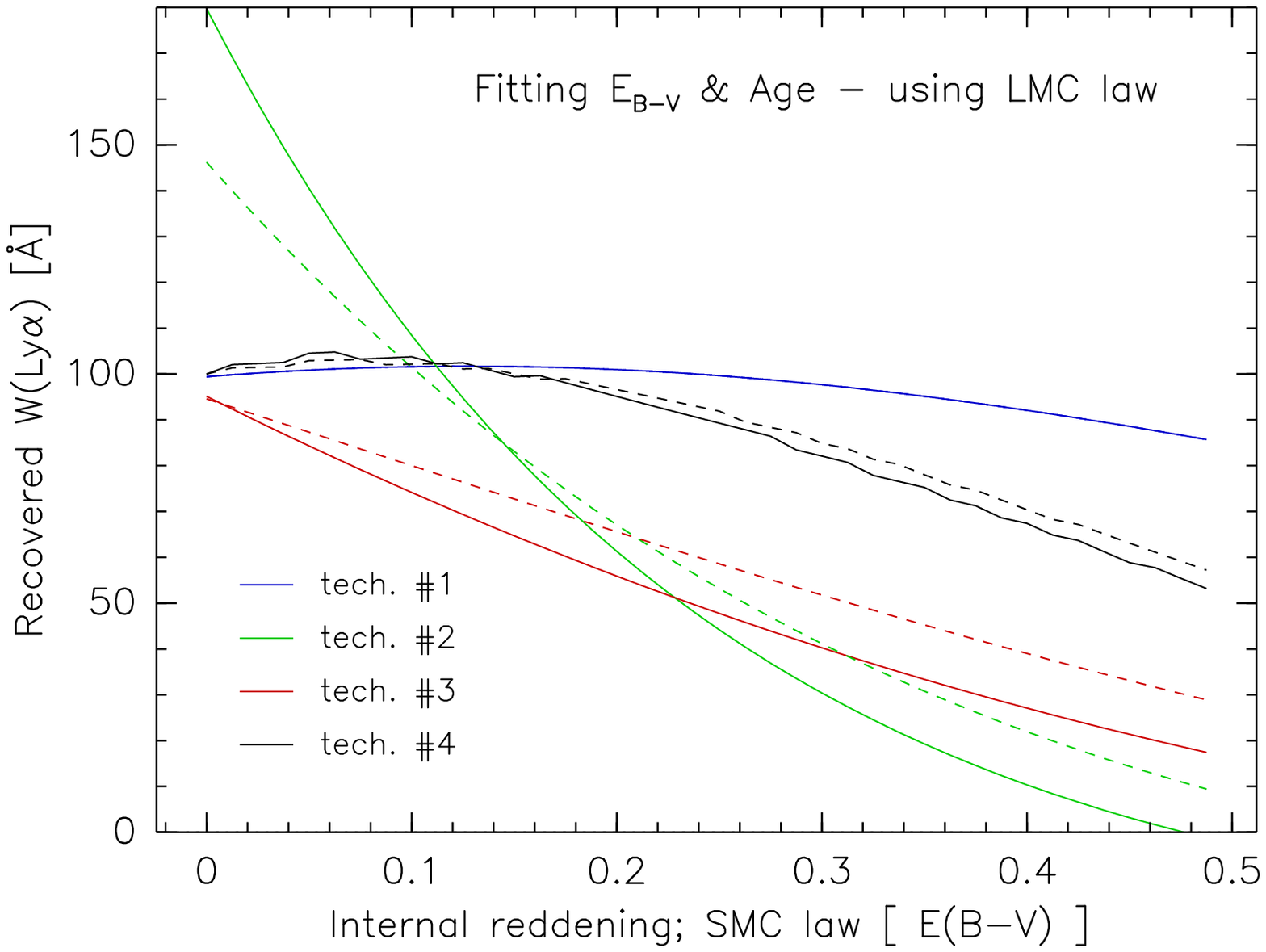}
\includegraphics[width=5.7cm,height=4.5cm]{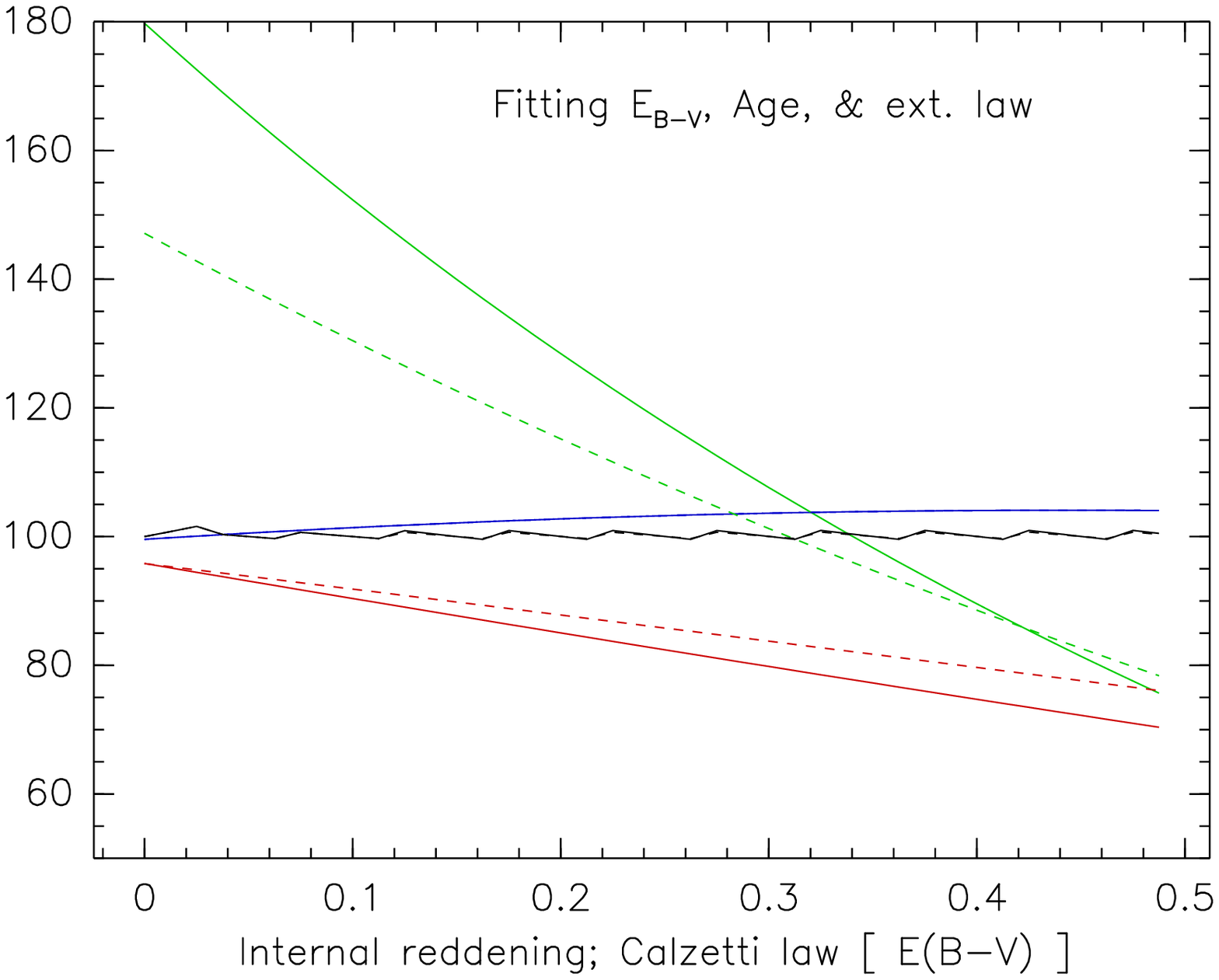}
\includegraphics[width=5.7cm,height=4.5cm]{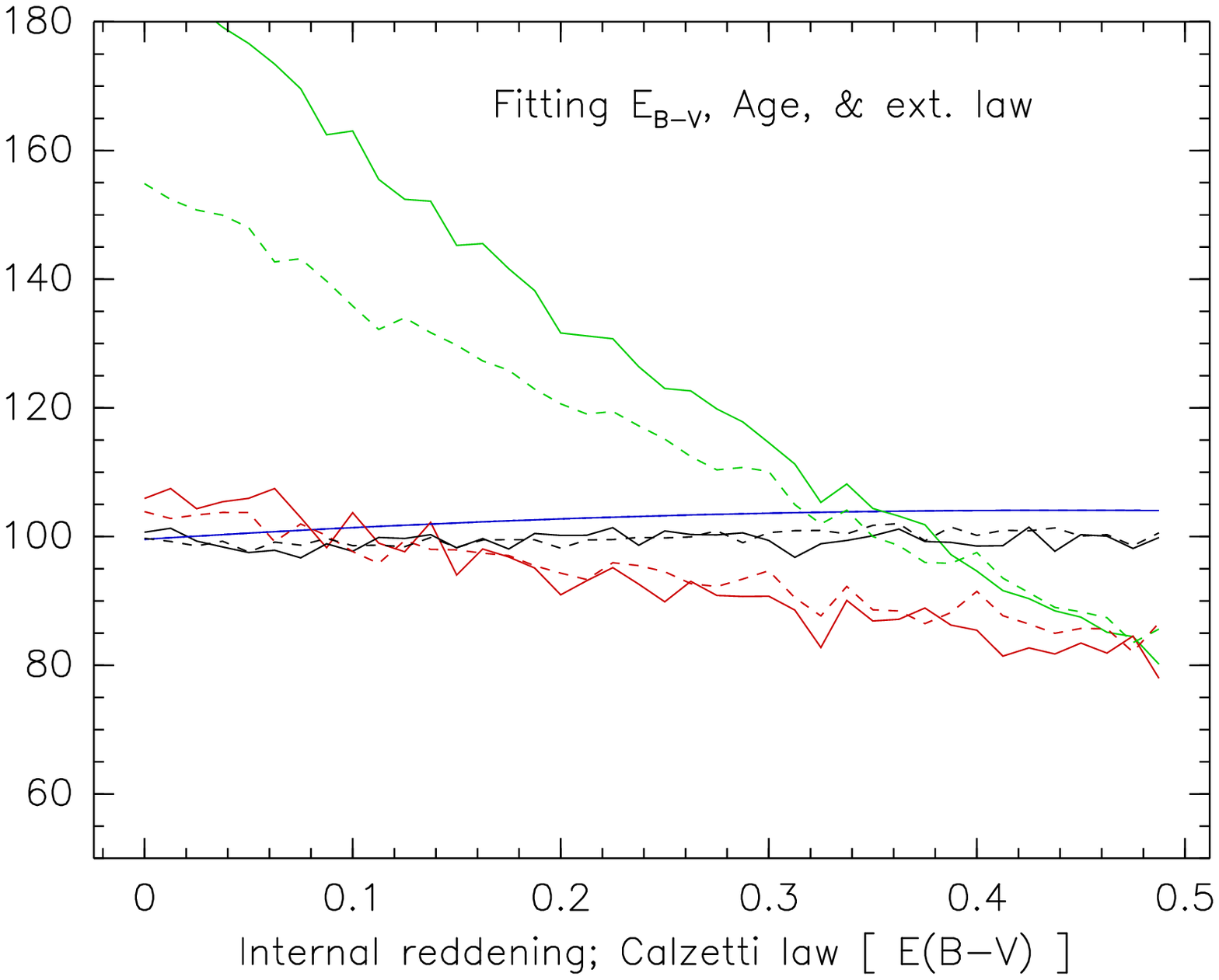}

\caption{Recovered \wlya\ as a function of internal reddening using the 
starburst extinction laws. 
The abscissa shows the internal reddening \ebv, using a known
extinction law.
The ordinate shows the recovered value of \wlya\ using an an extinction
law assumed by the observer. 
{\em Left}:  True dust: SMC-law; assumed dust: LMC.
{\em Centre}: True dust: Calzetti-law; extinction law free parameter in
SED fitting routine
{\em Right}: As for {\em centre} but with noise model applied.
Colours represent observational strategies as in Fig. \ref{fig:w_vs_lafwithz}. 
The filter setup is as described in section \ref{sect:fluxcomp} (solid lines). 
The dashed lines show the same results but with the nearest off-line
filter (restframe 1500\AA) shifted  to 1400\AA. 
[{\em See the electronic journal article for the colour version of this figure.}]
}
\label{fig:w_vs_ebv}
\end{figure*}

The tests presented here showed indistinguishable results at redshifts
of 2, 4, and 6, hence the results we present regarding reddening can be 
assumed to be independent of redshift. 
Fig. \ref{fig:w_vs_ebv} shows how well \wlya\ is recovered
for some test cases when dust is added to the system. 
The internal reddening for a given extinction law (type of dust) is 
represented on the abscissa of the
plots, while the recovered value of \wlya\ is presented on the ordinate. 
Where the SED-fitting technique (\#4) is concerned, the assumed
extinction law used by the SED-fitting routine may be different from that 
used to redden the restframe
SED and can be seen in the central caption of each figure. 
In the tests presented here we only redden the restframe SEDs using the
Calzetti and SMC laws because all other laws show a 2175\AA\ graphite
feature. 
If we believe we are dealing with objects with a very strong UV
radiation field, larger graphite-based molecules will be destroyed, 
removing any 2175\AA\ feature from the reddening vector and producing a
law similar to that of the SMC
(Mas-Hesse \& Kunth \cite{ref:mhkunth99}). 
We now only consider the impact upon the determination of \wlya, having
shown it to be a much more sensitive performance indicator than \flya.
Intervening \ion{H}{i} absorbing systems have been ``switched off" for
this experiment. 
The range in $0.0 < E_{B-V} < 0.5$ is chosen. 

Technique \#1 (blue line) is here demonstrated to very accurately recover
\wlya\ in all cases shown. 
At $E_{B-V}=0.5$ this technique overestimates \wlya\ by just 3\% using the 
Calzetti law (central panel) and the technique is largely insensitive to dust. 
This minor effect is a result of the steepening extinction curve removing flux 
from the broad bandpass, leading to a mild overestimate of the continuum
flux at \lya.
Conversely, this technique underestimates \wlya\ by 13\% for SMC-type
dust (left panel). 
Across the wavelength range of the 1216\AA-centred broad bandpass 
$(\sim 1000\AA$ to  $\sim1340\AA)$, the 
SMC-law provides more extinction than the Calzetti law for a given \ebv, 
and has a steeper gradient (more rapidly increasing with decreasing
wavelength). 
This results in the stronger suppression of the \lya\ line with the SMC law
than that of Calzetti, and turns the mild overestimate of \wlya\ into a
more significant underestimate. 


Technique \#2 (green lines) is highly ineffectual at reproducing the 
intrinsic 
equivalent width when dust is added to the system. 
This is also dependent upon the chosen extinction law: SMC-type dust
(left panel) 
completely suppresses the unreddened excess of a \wlya$=100$\AA\ line 
relative to the 1500\AA\ flux by $E_{B-V}=0.48$. 
That is, using technique \#2, at $E_{B-V}>0.48$ an LAE of
\wlya$=100$\AA\ will be seen as a \lya\ absorber. 
This effect is nowhere near as extreme when the Calzetti law is applied
since the SMC law is significantly steeper in the FUV. 


Modeling the UV continuum ($\beta$) as a simple power-law (technique
\#3; red lines) and extrapolating to \lya\ only performs marginally better in the 
presence of dust. 
A clear downward trend is visible with \ebv, although the
dependency is highly sensitive to the extinction law. 
At the relatively modest value of $E_{B-V}=0.1$, a \lya\ emitting object 
will have \wlya\ underestimated by around 10\% using the Calzetti law
but by 25\% for SMC dust.
By $E_{B-V}=0.5$ using the Calzetti law, \wlya\ is underestimated 30\%,
while for SMC dust the line has been almost completely suppressed. 
Extrapolating $\beta$ does not provide a reliable estimate of
\wlya\ when even a modest amount of dust is present since the dust
extinction curve modifies $\beta$ in such a way that it becomes
inconsistent with a power-law approximation. 
The dashed lines in Fig. \ref{fig:w_vs_ebv} show how recovery of \wlya\ is 
only very slightly improved when the continuum filter sampling 1500\AA\ is 
moved to 1400\AA.
Of course, it stands to reason that moving the off-line filter nearer to
the line is going to yield a better result and this is something that
must be considered when designing any observation -- even when the
filters can be ideally placed (which is also subject to the presence of
sky lines), the gain may not be significant.


Clearly technique \#1 is not greatly susceptible to the effects of internal
dust reddening.
In contrast, techniques \#2 and \#3 are highly susceptible to dust and
these observational configurations do not provide enough information to 
handle any reddening in a reliable manner.  
Continuum extrapolation techniques are not sufficient and a technique is
required that either handles the reddening explicitly or covers a
sufficiently narrow spectral domain. 
Indeed, the technique \#4 also becomes increasingly unreliable with
increasing dust content if the extinction law/dust type is not well
known -- assuming the wrong dust type renders the SED fitting method
ineffective (Fig \ref{fig:w_vs_ebv}, left panel). 
However, the 2200\AA\ continuum filter is well positioned to sample the
2175\AA\ graphite feature in the reddening vector.
This enables a third dimension to be added to the SED-fitting routine:
the discrete parameter of the dust extinction law. 
By fitting age, extinction law, and \ebv, we are able to recover
\wlya\ to within 1\% for all values of \ebv\ for each of the extinction
laws. 
See the central panel of Fig. \ref{fig:w_vs_ebv} for and example using
Calzetti law. 

Due to concerns about fitting three parameters with noisy data,
we adopted a similar noise-model to that described in Sect. 
\ref{sect:resdis:IGM}. 
Of particular concern was whether we could accurately recover the
extinction law. 
All fluxes were assigned $S/N=5$, randomised within the
corresponding Gaussian $\sigma$, and plugged back into the formulae to
compute \wlya. 
We now generate 1\,000 LAEs for each point in \ebv, and compute 
the mean values recovered. 
Since there are no \ion{H}{i} IGM clouds in this simulation, there is no
non-recovered population to that needs to be taken into consideration. 
The results can be seen in the right panel of Fig.
\ref{fig:w_vs_ebv}. 
Clearly, at the $5\sigma$ limit, the concerns about recovery of the
reliability of the fitting routine are not serious -- the SED fitting
routing recovers \wlya\ to within 3\% in all cases. 

Obtaining such observations of continuum-faint objects at high-$z$ requires a
substantial investment of time; many narrowband \lya\ 
surveys find a significant population of objects with no apparent UV or 
optical continuum.  
This is one of the areas in which the class of extremely large optical
telescopes has the opportunity to make a significant impact:  
with gains in collecting area of a factor of 25, restframe UV detections
of dusty systems may become more of a possibility. 
This is one of the reasons why we push this study to
seemingly extreme values of \ebv. 
If detections can be made, this will be of importance given that many 
\lya\ blobs (LABs) have been detected by narrowband imaging observations.
A significant fraction of these LABs are also bright sub-mm sources 
(Chapman et al. \cite{ref:champman05}; 
 Geach et al. \cite{ref:geach05}) 
implying a very significant dust
content and therefore reddening of the stellar continuum, although it
has not been demonstrated that the \lya\ and sub-mm radiation originate
from the same regions of the galaxy.
This is particularly true if the \lya\ production in LABs is the result
of accretion of cold gas onto dark matter halos
(eg. Haiman et al. \cite{ref:haiman00}) in which case the \lya\ may be
emitted over significantly extended areas. 
There may also be a certain degree of decoupling between
\lya\ photons and the nearby UV continuum due to resonance scattering. 
In the cases where there is no continuum detection at all in any bands, 
(eg. Nilsson et al. \cite{ref:nilsson06}) then equivalent width has no 
meaning. 
However, 
if there is a detection of (potentially reddened) stellar continuum then
total SED will be the superposition of this with the gas spectrum that
gives rise to \lya, whatever the mechanism for its production. 
In this case, the SED fitting approach should be applicable and \lya\ 
equivalent widths should be recoverable.

\subsection{Underlying stellar populations\label{sect:res:msp}}

Again intervening \ion{H}{i} absorbing systems have been switched off
for these experiments and we only consider the \lya\ equivalent 
width. 
Tests here are carried out using the standard template spectrum (defined
in Sect. \ref{sect:sim}, Tab. \ref{tab:stdpar})
combined with various old stellar populations at 
redshifts of 2, 4, and 6. 
However, within the tests performed here, no redshift dependence at all
was detected and hence, none will be discussed. 
The results presented here can be considered to hold at all redshifts.

\begin{figure}[htbp]
\centering
\includegraphics[width=7.0cm]{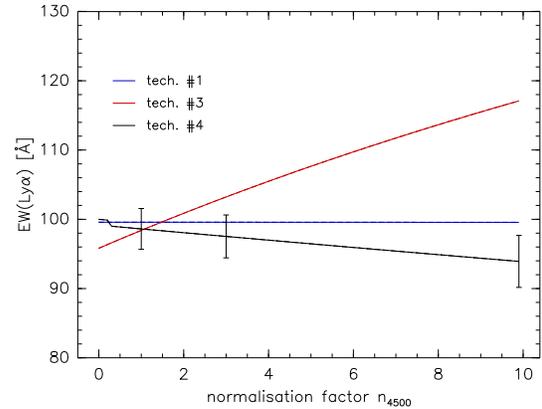}
\caption{The effect of varying $n_{4500}$ on the recovered values of
\wlya\ as determined by the various techniques. 
$n_{4500}$ is defined in Sect. \ref{sect:msp}.
The error bars on the black line show the standard deviation in recovered \wlya\
when 1\,000 objects
are generated using the noise model described in the text. 
[{\em See the electronic journal article for the colour version of this figure.}]
}
\label{fig:osp}
\end{figure}

Tab. \ref{tab:mspset} shows the way in which varying the contribution
from an old stellar population affects the computed values of \wlya\
when the age and metallicity of the underlying population are
varied. 
Fig. \ref{fig:osp} demonstrates the effect of increasing the
relative contribution of a 300Myr underlying population.

\begin{table}[htbp]
\caption{The influence of old stellar populations on detected \wlya\ -- the
effects of age and metallicity.} 
\label{tab:mspset}
\flushleft
\begin{tabular}{c c c c c c c}        
\hline\hline                 
Age   & $Z$ & $n_{4500}$ &   \multicolumn{4}{c}{\wlya\ from technique}\\ \cline{4-7}
(Myr) &     &  &   \#1 & \#2 & \#3 & \#4\\
\hline                        
\hline                        
-- & --  & 0.0 & 99.581 & 179.774 & 95.819 & 100.000 \\

\hline
900 & 0.008 & 1.0 & 99.579 & 178.865 &  98.398 &98.610 \\
600 & 0.008 & 1.0 & 99.524 & 175.017 &  96.486 &96.282 \\
300 & 0.008 & 1.0 & 99.530 & 168.273 &  92.709 &92.202 \\
200 & 0.008 & 1.0 & 99.552 & 165.828 &  91.587 &90.723 \\
100 & 0.008 & 1.0 & 99.557 & 164.678 &  90.934 &90.027\\
\hline
300 & 0.001 & 1.0 & 99.511 & 171.039 &94.156 &93.875 \\
300 & 0.004 & 1.0 & 99.539 & 168.296 &92.505 &92.216 \\
300 & 0.020 & 1.0 & 99.478 & 171.927 &94.899 &94.412 \\
300 & 0.040 & 1.0 & 99.550 & 176.421 &97.581 &97.131 \\
\hline                       
\end{tabular}
\flushleft
Note: Normalisation factor ($n_{4500}$) defined in Sect. \ref{sect:msp}.\\
\end{table}
For the tests presented for the age and metallicity of the underlying 
population, $n_{4500}$ is set to 1 in every case, meaning the underlying
population and starburst SEDs contribute equally in the $B$-band. 
Clearly technique \#1 is entirely unaffected by the presence of an aged
stellar population, being in error by no more than 0.5\% in every case. 
Again, the strength of this technique results from the fact it is dependent
only on a very narrow spectral region. 
Technique \#2 is consistently inaccurate, showing the typical
overestimate of \wlya\ by $\sim70$\% but is self-consistent to within
about 10\%. 
Technique \#3 is accurate to within 10\%. 
The 4-filter SED-fitting technique (\#4) implemented here is included in
this study 
because we have demonstrated it to be a reliable technique by which the
continuum at \lya\ can be accurately estimated (Hayes et al.
\cite{ref:hayes05}).
The strength of this technique relies on the UV
continuum slope and 4000\AA\ break being adequately sampled: then the effects 
of age and reddening aren't completely degenerate and can be disentangled. 
Naturally, if the 4000\AA\ break is so important, it is possible that
the presence of an additional stellar population that contributes
strongly to the
4000\AA\ break but not in the UV could render this technique ineffective.
No real trend of \wlya\ with metallicity is observed with any of the
techniques. 
A trend of \wlya\ with the age of the underlying stellar
population does emerge when computed using techniques \#2, \#3,
and \#4: recovered \wlya\ decreases as the underlying population gets younger. 
The populations are normalised in the restframe at 4500\AA\ but the
SEDs become bluer with decreasing age.
Hence, applying a younger underlying population has the effect of making
the composite SED bluer than when an older population is added. 
As a result, the younger the underlying population, the higher the flux
at 1500\AA\ which leads to a tendency to overestimate the continuum at
\lya. 
Therefore the estimates of the continuum level at \lya\ become higher
and the computed equivalent width is decreased.

When considering dependencies upon the normalisation coefficient, no
trends emerge when using technique \#1, (for the same reasons this
technique is not sensitive to the age of the underlying population; 
Fig. \ref{fig:osp}). 
Technique \#3, which extrapolates the measured value of \bet\ to \lya,
tends to overestimate \wlya\ with increasing $n_{4500}$.
As $n_{4500}$ increases, the flatter SED of the old population has the
effect of flattening in the spectrum in the 1500-2200\AA\ region. 
However, the 300Myr burst turns over towards \lya\ much more sharply than 
the very young SED, and hence doesn't contributed much at 1216\AA. 
Hence the continuum flux at \lya\ is increasingly underestimated as
$n_{4500}$ increases.
There is a minor tendency for technique \#4 to become less accurate with
increasing $n_{4500}$, underestimating \wlya\ by 6\% at $n_{4500} = 10$. 
The
fitting software selects increasingly older template spectra as the old
population contributes more and more to the 4000\AA\ break. 
However, with increasing $n_{4500}$, the composite spectrum changes
faster across the 4000\AA\ break than it does in the FUV.
So as older (redder intrinsically) spectra are selected by the fitting 
software, in order to maintain a good fit in the UV,
the best-fit SED will be {\em less} reddened than it would otherwise.
This leads to a tendency to slightly overestimate the continuum flux at
\lya, and a trend of underestimating \wlya\ with increasing contribution
from an older population. 

In light of these results we made a number of attempts to confuse our
SED-fitting software. 
The first was to, as in previous sections, see how the SED fitting
routine fared when the simple noise model was applied.
After applying the aged population at each value of $n_{4500}$, we 
randomised all the computed fluxes by assigning $S/N=5$ in each filter, 
and computed \wlya\ from these values. 
We generated 1\,000 objects and computed the mean values although these
were never found to be deviant from the values computed without the
noise model at all $n_{4500}$.
The standard deviation of the recovered \wlya is represented by the 
error-bars on the black line (technique \#4) of Fig. \ref{fig:osp}.

Secondly, we randomly selected old populations from any of the templates with
ages greater than 300Myr,
reddened them with random \ebv\ in the range 0.0 -- 1.0, and
added them to the template SED with varying $n_{4500}$ (range: 0.0 -- 3.0).
By applying two old populations to our template spectra in this fashion, we 
were not able to make technique \#4 reproduce \wlya\ that was in error by 
more than 10\%. 

The final test devised to confuse the software was to vary the
contribution of the nebular continuum in the template spectra. 
The motivation for this test again being that the SED-fitting software is
sensitive to the Balmer jump.
Moreover, our understanding of the ionising
photon production from massive stars is incomplete.
There is a lack of data concerning low-metallicity stars, no single 
O type stars can be observed at such FUV wavelengths, and the ionising
contribution from population {\sc iii} objects is largely speculative.
Additionally, Lyman-continuum escape has been observed in very few 
cases: 
Bergvall et al. (\cite{ref:bergvall06}) 
at low-$z$ in the case of ESO\,350-IG38; and 
Steidel et al. (\cite{ref:steidel01})
at high-$z$ by stacking composite spectra of Lyman Break Galaxies (LBGs). 
It is the reprocessing of these photons that determines the intensity 
of the nebular component relative to that of the stars, hence any
Lyman continuum photons that escape are not reprocessed in the 
nebular hydrogen emission spectrum (lines or continuum).
We found that, even when scaling the nebular component (see Sect.
\ref{sect:sedgen}) by a factors in the range $0-3$, technique \#4
recovered \wlya\ with errors no greater than 5\%. 
These limits extend between two very extreme cases: one, where all Lyman 
continuum photons escape (which would result in no nebular emission
component and therefore no \lya); and two, the possibility that the estimates 
of ionising continuum in stellar atmosphere models are in error 
by a factor of 3 and all the ionising photons are reprocessed as 
nebular emission.

\subsection{Redshift and observational dependencies}

One of the initial motivating factors for the study of \lya, dating back to
Partridge \& Peebles (\cite{ref:pp67}), was the fact that it could
provide a tracer of early star formation observable from the ground at
redshifts greater than around 2. 
At $z=2$, technique \#1 would be equivalent to a narrowband filter at
3600\AA, plus a $U$ filter. 
Here the effect of intervening \ion{H}{i} systems will be
lower than at higher $z$ and, at the lowest redshifts where \lya\ is
observable from the ground, intervening \ion{H}{i} affects the
determination of \wlya\ by technique \#1 at the 1\% level for our chosen filter
set.
That said, any single measurement by this technique cannot be deemed
reliable without additional observations. 
Followup spectroscopy could obviously confirm the presence of
intervening systems and would also yield an accurate measurement of
\wlya, but observational requirements may preclude the possibility of
obtaining such data for every target. 
Perhaps cheaper, depending on the design of the survey, would be to use
two off-line filters, say $B$ and $V$ (restframe 1470\AA\ and 1830\AA,
respectively -- where CCDs are typically more sensitive than in $U$), in
order to estimate the continuum level at \lya\ without sampling
bluewards of 1216\AA. 
However, the 
Shapley et al. (\cite{ref:shapley03})
LBG sample find median \ebv\ of 0.099 for the strongest \lya\ emitters
with \ebv\ increasing with decreasing \wlya. 
At $E_{B-V}=0.1$, technique \#3 is already underestimating
\wlya\ by 25\% for SMC dust. 
The dust-sensitivity of technique \#3 is independent of redshift and
will not be discussed further. 
At $z=2$, the 4000\AA\ break can still be sampled by an observation in
the $J-$band allowing technique \#4 to be used in order to properly
handle the effects of dust reddening. 

As redshift increases, the effect of intervening \ion{H}{i} on technique
\#1 becomes more pronounced, reaching the 10\% level by $z=4$. 
Systematic uncertainties on numbers derived by this technique only should
probably be addressed using a Monte Carlo-type approach. 
At this redshift, the 4000\AA\ break can still be sampled from the
ground using the $K_\mathrm{s}$ band although the high night-sky background
makes these observations expensive. 
In 40 hours of integration time on an 8 metre telescope, 
a typical $K_\mathrm{s}-$band imaging camera can detect an object of 
$K_\mathrm{s}=24$ in the $AB$ magnitude system at $S/N = 5$ 
\footnote{Computations are performed with the Exposure Time Calculator for the
{\em Infrared Spectrometer And Array Camera} (ISAAC) mounted on 
{\em Very Large Telescope} (VLT) at ESO Paranal, in ideal observing conditions
(seeing of 0.7\arcsec; airmass of 1.2).}.
At $z=4$, and assuming the local $B-$band luminosity function 
(Jones et al. \cite{ref:jones06})
evolves parallel to the LF at 1500\AA\ between redshifts of 0, 
(Wyder et al. \cite{ref:wyder05})
and 4
(Yoshida et al. \cite{ref:yoshida06}), 
this corresponds to a detection limit 0.5 mag fainter than $M^\star$.
According to {\em James Webb Space Telescope (JWST)} Mission Simulator and 
NIRCam sensitivity estimates,
such observations to these depths can be obtained at $2\mu m$ in just 1 second. 

At redshifts greater than 5, $L-$band observations would be required in
order to sample the 4000\AA\ break.
While deep observations may be obtainable from the ground at shorter 
wavelengths, the sky-background at wavelengths longer than 
$K_\mathrm{s}$ preclude such observations and current groundbased
facilities do not approach the required sensitivities. 
Consequently these observations would need to be carried out by
mid-infrared telescopes in more unfriendly and expensive environments:
for example the Spitzer Space Telescope or JWST, or a proposed MIR telescope 
at Dome-C, Antarctica.
This is somewhat contrary to the motivation for using \lya\ as a probe
for primeval star-formation. 
However, we reiterate that it is only this observation that needs to be
done from space -- the other bands may, depending on redshift,  more 
economically be performed from the ground. 
Moreover, in the age of deep multi-band surveys, the amount of
pre-existing data there is for a field is always a consideration. 
If observing a deep field at a certain wavelength, it may well be
advantageous to select fields for which much data is already in existence
-- observations at new wavelengths are frequently added to existing
datasets (the Chandra Deep Field, for example).
Hence the majority of the observations required for technique \#4 
(targeting certain redshifts, at least) may already be in place, requiring 
only the additional infrared bands and  it would be wise for future 
\lya\ surveys to capitalise on the currently existing data.
It is also likely that early JWST programs will include deep observations of
regions with existing deep optical data.

In addition to the observational, there are also further theoretical
uncertainties; mainly concerning the validity of the stellar atmosphere 
models in the UV. 
These models have currently not been well tested at these wavelengths. 
See Sect. \ref{sect:res:msp} for a discussion on this. 



%
\section{Conclusions and summary\label{sect:conc}}
Simulations of observations of high-redshift \lya\ emitting galaxies
have been performed.
Specifically, the simulations have examined how efficiently the 
intrinsic values of \lya\ flux and equivalent width are recovered by 
narrowband imaging observations. 
\flya\ and \wlya\ are determined by the four methods described in Sect.
\ref{sect:sim} and the effects of intervening \ion{H}{i} clouds, 
internal dust-reddening, and underlying stellar populations, 
have been investigated. 
In summary: 
\\
-- Observing \lya\ with one off-line continuum filter has been
shown to be highly ineffectual. 
The steep UV continuum slope ($\beta$) may cause overestimates of 
\wlya\ by a factor of almost 2, while the neglection of dust reddening 
may cause the reverse effect: narrowband imaging observations may 
interpret bright \lya\ emitters (\wlya$>100$\AA) as \lya\ absorbing 
systems. 
\\
-- Using two filters to estimate the UV continuum slope and
extrapolating to \lya\ using a power-law improves the situation to
varying degrees.
In the dust-free cases this technique becomes a highly competitive method
of determining \flya\ and \wlya.
However, even modest amounts of dust render this technique highly
ineffectual since dust modifies the continuum in such a way that the
power-law parameterisation becomes unreliable. 
$E_{B-V} = 0.1$ results in 25\% underestimate of \wlya.
\\
-- Economic observing strategies that utilise a narrowband and 
single broadband filter with the same central wavelength are far less
susceptible to dust. 
However, the continuum flux estimation in such observations becomes 
suspect due to the blue half of the broadband filter being on the blue
side of \lya, and hence being susceptible to absorption of the continuum 
by intervening \ion{H}{i} clouds. 
Such a technique can lead to overestimates of \wlya\ by factors of
$>2$. 
With the addition of a simple sky-noise model, this effect becomes
more pronounced with decreasing $S/N$.
We suggest that in such observations, simulations are performed to
estimate the biases caused by such effects.
\\ 
-- SED-fitting techniques that observe only redwards of \lya\ are not
susceptible to intervening \ion{H}{i} absorption or dust reddening,
when a single extinction law is considered at all wavelengths.
Such techniques need to sample the UV continuum slope, 2175\AA\ dust feature,
and 4000\AA\ break.
This way reliable estimates of the continuum flux at \lya\ can be made and
\wlya\ estimates are much better constrained in the presence of noise.
\\
-- None of the techniques we considered here were hugely sensitive to 
the presence of underlying old stellar populations or randomly mixed
populations. 
Additional populations have no detrimental effect on the efficiency of
the technique utilising two filters centred on \lya.
The contribution of such populations to the 4000\AA\ break cause only
minor ($<10$\%) errors in the SED-fitting routine. 
Similarly increasing the relative contribution of nebular emission 
has no significant impact upon the estimates of \flya\ and \wlya. 
\\
-- Independent of technique, we find a redshift-dependent incompleteness
that results from \ion{H}{i} systems along the line-of-sight.
Such an effect would have to be allowed for in the compilation of a \lya\
luminostiy function.
The red damping wing of DLAs close to the LAE can remove sufficient flux
to completely suppress the \lya\ line. 
At $z=6$, one could expect to miss around 10\% of the sources using
our perfect filter set.

We have shown that to recover \flya\ and \wlya\ from high-$z$ LAE 
systems, a SED fitting method is preferable if observations can reach
the required depths.  
Other techniques suffer badly from either \lya\ absorption along the
line-of-sight or reddening by internal dust. 
If such off-band observations are not available, the use of a single narrow-
and broadband filter with the same central wavelength seems to be
preferable, although such observations should be complemented with some
tuned statistical study of the effects of intervening \ion{H}{i}
systems.

\begin{acknowledgements}

We acknowledge the support of the Swedish National Space Board
(SNSB) and the Swedish Research Council (VR).
We thank J.~M. Mas-Hesse, D. Kunth, J.~P.~U. Fynbo for their valuable 
comments on this manuscript and C. Leitherer and A. Petrosian for 
their work on the Lyman-alpha projects.
We would also like to thank the anonymous referee for comments that have sparked
numerous improvements to the manuscript. 

\end{acknowledgements}

\end{document}